# Title

Brain MRI Segmentation using Template-Based Training and Visual Perception Augmentation

# Author list

Fang-Cheng Yeh[1,2,*]

# Affiliations

[1]Department of Neurological Surgery,

University of Pittsburgh, Pittsburgh, Pennsylvania, USA

[2]Department of Bioengineering,

University of Pittsburgh, Pittsburgh, Pennsylvania, USA

*Corresponding author: frank.yeh@pitt.edu

Fang-Cheng (Frank) Yeh

Associate Professor of Neurological Surgery

University of Pittsburgh School Medicine

3550 Terrace Street, Scaife A507

Pittsburgh, PA 15261

## Abstract

Deep learning models usually require sufficient training data to achieve high accuracy, but obtaining labeled data can be time-consuming and labor-intensive. Here we introduce a template-based training method to train a 3D U-Net model from scratch using only one population-averaged brain MRI template and its associated segmentation label. The process incorporated visual perception augmentation to enhance the model's robustness in handling diverse image inputs and mitigating overfitting. Leveraging this approach, we trained 3D U-Net models for mouse, rat, marmoset, rhesus, and human brain MRI to achieve segmentation tasks such as skull-stripping, brain segmentation, and tissue probability mapping. This tool effectively addresses the limited availability of training data and holds significant potential for expanding deep learning applications in image analysis, providing researchers with a unified solution to train deep neural networks with only one image sample.

## Introduction

The difficulty of obtaining sufficient expert-labeled data has profoundly impacted deep learning(Keshari et al., 2020; Thomas et al., 2020). Deep learning models rely on enough training data to effectively learn and generalize to new situations (LeCun et al., 2015; Zhang et al., 2021), but getting labeled data is a time-consuming and labor-intensive process. In application domains such as biomedical imaging, labeling or segmenting the image may require domain expertise and meticulous manual delineation by trained professionals. These challenges, coupled with the exponential growth of data, make it increasingly difficult to gather a sufficiently diverse and comprehensive set of labeled images. Without sufficient expert-labeled data, the performance of these models may be limited, and their ability to accurately process and analyze images may be hindered(Shen et al., 2017). This can impede imaging AI's progress and development and its biomedical research applications.

One commonly-used approach to overcome the challenge of limited data in imaging AI is *data augmentation* through geometry transforms such as padding, random rotating, re-scaling, vertical and horizontal flipping, translation, cutout, and zooming (DeVries and Taylor, 2017; Krizhevsky et al., 2017; Yang et al., 2022). Other approaches include photometric transforms such as darkening & brightening/color modification. These approaches have been commonly applied to augment the training data but have not yet shown a capability to train a deep neural network on very few samples. In contrast, human visual processing possesses a remarkable ability to learn and generalize from a single template. Neuroanatomists, for instance, can learn and accurately identify brain structures from a single neuroanatomy textbook, highlighting the remarkable adaptability of human visual perception to learn from a single template. This insight inspired us to explore solutions that enable training a deep neural network using just a single template, akin to the role of a neuroanatomist.

In this study, we introduce a novel approach that leverages publicly-available brain image templates and their corresponding tissue segmentation labels (Ahmad et al., 2023; Ciric et al., 2022) to train U-Net models from scratch (Fig.1). Our methodology enabled the training of U-Net models to generate brain skull-stripped images, voxel-wise segmentation labels, and tissue probability maps, for commonly studied species in neuroscience, including mice, rats, marmosets, rhesus monkeys, and humans.

Our training approach began by gathering population-averaged templates and tissue segmentation data for commonly studied species in neuroscience, such as rats, mice, marmosets, rhesus monkeys, and humans (Suppl. Table 1). We thoroughly examined the segmentation data to ensure consistency across the labels, which include white matter, gray matter, cerebellar cortex, basal ganglia, and ventricles. Subsequently, we introduced a novel image augmentation technique that mimics the adaptability of human visual perception to diverse visual input conditions (Thompson et al., 2011). The augmentation steps included image reduction, noise and cropout introduction, simulation of various lighting conditions, rigid body and camera transformations, and addition of background textures (Fig. 1b and 1c). The reduced resolution, added noise, missing parts, and varying illuminations created a training environment that simulated the diversity of visual inputs encountered in real-world scenarios(Thompson et al., 2011). Furthermore, we addressed the challenge of viewpoint dependence problem(Tarr and Pinker, 1989; Tarr et al., 1998) by *limiting* the parameters used in the rigid body and camera transformation. Unlike conventional image augmentation techniques that imposed image flipping and large, unlimited rotations(DeVries and Taylor, 2017; Krizhevsky et al., 2017; Yang et al., 2022), we carefully chose the upper limit of viewport parameters to mitigate their adverse effects on object recognition. Moreover, the artificial textures introduced into the background allowed the trained network to recognize objects even when partially occluded or presented in challenging imaging settings. This background augmentation aligned with the figure-ground perception process in human visual perception(Wagemans et al., 2012), where objects are distinguished as separate entities from their background. Overall, the entire *visual perception augmentation* framework synergistically combined vision-based augmentations, aligning them with human visual perception capabilities. The augmentation provided deep neural networks with a deeper understanding of the template input, enabling them to excel in diverse and challenging environments using only a single template for training.

After training the models, we extensively evaluated the capability of our models across different species and their performance in scenarios involving brain lesions. The visual perception augmentation process and the trained models for multiple species were integrated into an open-source and user-friendly tool, providing researchers with a seamless and efficient resource for demanding brain segmentation tasks in both human and animal brain MRI.

# Materials and Methods

### *U-Net architecture*

The 3D U-Net was implemented using C++ (MSVC2019) and libtorch (version 1.13.0+cu117). The U-Net architecture used in this study resembles the original U-Net Study(Ronneberger et al., 2015), with changes in the number of features and the use of 3D convolutional and upsampling layers. As shown in Fig.1a, the architecture consists of 5 layers, with feature numbers 8, 16, 32, 64, and 128. Each encoding layer has two 3-by-3 kernel convolutional layers, ReLU activation, and batch normalization, followed by max pooling to downsize the images by a factor of 2. In the decoding layers, the input combines the last images of the encoding layer with the upsampled image from the lower layer. The decoding layers include two 3D convolutional layers, ReLU activation, batch normalization, and an upsampling layer (except for the final decoding layer). The final layer is a 3D convolutional layer with a one-by-one kernel. We used mean squared error as the loss function for backpropagation. The optimization was done using the Adam optimizer.

### *Visual perception augmentation*

The augmentation steps were implemented using the C++ programming language with source code available to reproduce the same results (https://github.com/frankyeh/UNet-Studio/). Before augmentation, the only preprocessing step of the input template image was scaling the values by a constant such that the maximum value was one. This scaling was conducted again after the augmentation so that the augmentation image also ranged between 0 and 1. The visual perception augmentation process (Fig. 1b and 1c) involves several steps described in the following sections:

*Subsampling*

The subsampling procedure first reduced the template image's width, height, and depth to half of the original ones. This size reduction was applied at 50% chance at each dimension independently. After volume reduction, the image was upsampled back to the original size using linear interpolation. The entire process introduced a certain degree of blurring to the resulting image, a consequence of the interpolation method employed during upsampling, which tends to smooth out the fine details and high-frequency information in the original image.

*Uniform noise*

The uniform noise was added independently to each voxel using a uniformly random value between 0 and 0.2. This intentional addition of random values resulted in the image acquiring a visually noisy appearance. The uniform noise contributes to the introduction of variations and fluctuations across the image, effectively simulating the presence of noise and enhancing the overall realism of the image.

*Cropping*

The cropping step removed a spherical area from template images and their associated labels. The cropped area in the template image was replaced with a constant value randomly selected between 0 and 2, whereas the cropped region in the label image was replaced by 0. The radius of the cropped area was uniformly randomly determined between 0.1 and 0.2 of the image width. Additionally, the center of the spherical area was also uniformly randomly selected. In this study, cropping was only applied with training the human model for brain tumor images. The rest models were trained without cropping.

*Ambient, diffusion, and specular light*

The ambient light was implemented by adding a constant value to the image. This value was uniformly and randomly chosen between 0 and 2 and was identically added to every voxel in the image volume, creating a uniform illumination. Then, the diffuse light was added by first generating a randomly oriented 3D unit vector. This unit vector represented the direction of the light source. Then, for each voxel, an inner product was computed between the unit vector and the displacement vector between the voxel and the image center divided by the image width. The voxel intensity was multiplied by the inner product multiplied by 0.2. This resulted in variations in brightness across the image based on the direction of the unit vector. Last, specular light was introduced by first randomly determining a specular center within the image. Each voxel's distance to the specular center was calculated and divided by the image width. Subsequently, the voxel's intensity was multiplied by the scaled distance. This process produced varying levels of brightness and darkness across the image, depending on the distance between each voxel and the specular center. By incorporating these lighting effects, the resulting image exhibited realistic variations in brightness and illumination, enhancing its visual appearance and depth.

*Rigid motion and camera transform*

The implementation of rigid motion and camera transforms involved a coordinate conversion function that converted a 3D coordinate vector from the destination coordinates $\vec{u}$ back to the source coordinates $\vec{v}$. The rigid motion encompassed rotation and translocation, while the camera transforms encompassed scaling, aspect ratio adjustment, perspective transformation, and lens distortion. These transformations were applied backward, starting with lens distortion and proceeding with perspective transformation and other remaining effects. The destination vector was converted back to the source image space by performing the transformations in reverse, enabling cubic spline interpolation for image volume transformation.

To simulate lens distortion, a distortion field, $d(\vec{u})$, was first added to the destination coordinates: $\vec{u} + d(\vec{u})$. This distortion field was implemented using a Taylor expansion estimation of the pincushion distortion(Vass and Perlaki, 2003). The distorted distance $d(\vec{u})$ was defined as $d(\vec{u}) = m(\vec{c} - \vec{u})||(\vec{c} - \vec{u})||^2$, where $\vec{c}$ represents the center coordinate of the image volume, and *m* is a constant that controls the magnitude of the distortion. The value of *m* was calculated by multiplying half of the maximum image width by a randomly chosen constant between 0 and 0.1. This constant remained the same for all imaging voxels. The distortion field was added to the destination coordinates to introduce lens distortion. As a result, the image exhibited varying degrees of lens distortion, creating a visually distinct effect characterized by non-uniform stretching and warping.

After applying the lens distortion field, the perspective transformation was applied by scaling the coordinates using the formula: $\frac{\vec{u}}{\vec{p}(\vec{u}-\vec{c})+1}$. In this equation, $\vec{u}$ represents the coordinates after lens distortion, $\vec{c}$ represents the center coordinate of the image volume, and $\vec{p}$ is a perspective vector that determines the magnitude of the perspective transform. The perspective vector $\vec{\mathrm{p}}$ is derived from three randomly chosen constants between -0.5 and 0.5. Each constant was divided by the image width in the x, y, and z dimensions.

After applying lens distortion and perspective transformation, the remaining camera transform, including scaling and aspect ratio adjustments, as well as the rigid motion involving rotation and translation, were implemented using a simple linear equation: $\vec{\mathrm{v}} = \mathbf{R}\vec{\mathrm{u}} + \vec{t}$. In this equation, **R** is a 3-by-3 matrix incorporating rotation, scaling, and aspect ratio adjustments. The rotation values were randomly chosen constants between 0 and 0.2 rad, representing the desired rotational effects. The translocation vector $\vec{t}$ was determined by three random constants between 0 and 0.2, multiplied by the image sizes at each dimension. This vector defines the shift in the position of the transformed coordinates. Additionally, the scaling factor was determined by a random constant between 0.8 and 1.25, allowing for variable scaling of the image. The aspect ratio was determined by a random constant between 1.0 and 1.25, enabling adjustment of the image's width-to-height ratio. The image undergoes the combined effects of scaling, aspect ratio modification, rotation, and translation by applying the linear equation, resulting in the desired transformation.

These transformation parameters were determined and tested based on the first author's prior training in neuroanatomy: the testing examined the maximum transformation values that still allowed for recognizing brain structures.

*Background textures*

We utilized two artificial background textures: stamping and Perlin noise texture(Perlin, 1985). The background and foreground regions of the template were determined based on the label information. The background region had a zero label, while the foreground region had non-zero labels. Each background texture was applied with a 50% probability. The stamping texture involved repeatedly drawing the rotated and scaled background image (excluding the foreground) onto the background. All possible rotation angles were utilized, and the translation was limited to a maximum of half the image width. The scaling factor was randomly chosen between 0.8 and 1.25. The stamping process was repeated five times to populate the background with its texture pattern, while the foreground voxels remained unchanged. The stamped background was blended with the original background using the blending function represented by the formula: $s \leftarrow s + bf(s)$. In this formula, $b$ represents the newly added intensity, and f is a scaling function. The scaling function is defined as follows: if 1 - s is less than 0.1, the function outputs 0.1; otherwise, it outputs 1 - *s*. The 3D Perlin texture was created by applying a floor function to the 3D Perlin noise(Perlin, 1985). The same blending function drew the Perlin noise texture to the background.

***Template-based training***

The template-based training was conducted in three rounds with different learning rates. The first round had 1,000 epochs. At each epoch, 8 augmented images were generated from visual perception augmentation. Each augmented image was forwarded through the U-Net at a time to reduce GPU memory requirements. The loss was calculated, with subsequent backpropagation conducted. Finally, after accumulating gradients from 8 augmented images, the epoch was ended by updating the model using an Adam optimizer with a learning rate of 0.001. The accumulated gradients were equivalent to a batch size of 8 training images. The second round was then conducted with a learning rate of 0.0005, 2,000 epochs, and 16 augmented images for each epoch. Finally, the third round was completed with a learning rate of 0.00025, 2,000 epochs, and 32 augmented images for each epoch.

Two errors were reported in each epoch throughout the training process. The first type was the training error calculated based on the loss function used for backpropagation, providing insights into the model's optimization progress. The second type of error was obtained by evaluating the original template and termed the template error. It is worth emphasizing that the original template was never used during training, and no backpropagation was performed based on it.

All models were trained from scratch using the aforementioned template-based training approach. The training was conducted on a Dell Precision 7820 Tower equipped with two Intel Xeon Silver 4114 CPUs and an NVIDIA RTX A6000 48GB Graphics Card. The computation time for each model varied depending on the image size and typically took around a day.

***Training and evaluation data***

The sources of the training templates and evaluation data are listed in Suppl. Table 1 and 2, respectively. Suppl. Table 1 lists each template and label used in training the model. Since different templates have segmentation labels, we merged or separated the existing segmentations into five tissue/region types: (1) white matter (including brainstem and cerebellum white matter), (2) gray matter (including amygdala and hippocampus), (3) cerebellar cortex, (4) basal ganglia (including putamen, globus pallidus, caudate, thalamus), and (5) ventricular spaces, subarachnoid space, and choroid plexus. The first author conducted a slice-by-slice quality check based on experiences in priors studies of neuroanatomy.

***Preprocessing and post-processing steps***

The only preprocessing step used was scaling the image value so that the maximum equaled one. In this study, we did not utilize bias field correction to correct the image inhomogeneity. If the evaluation images have a different resolution from the templates, the images will be resampled using cubic spline interpolation to match the spatial resolution of the training templates.

The U-Net model produced a 4D tensor that contains five 3D volumes, each corresponding to one of the five tissue segmentations. The output values were first thresholded between 0 and 1. A brain mask was then constructed by summing the 4D tensor output into one 3D volume, binarizing it with a threshold of 0.5, and identifying the largest region using a morphological operator with a 6-neighbor connection. The constructed mask was then used to zero the background of the input image to produce the skull-stripping results. On the other hand, the mask was used to zeros the five 3D volumes as the tissue probability maps. The 3D label was then generated by searching for tissue with the maximum value among the tissue probability maps.

### *Comparason with other brain segmentation tools*

The evaluation image employed in this comparison was the T1-weighted image of the HCP-YA subject #100206, which was resampled to a resolution of 1-mm. The FastSurfer(Henschel et al., 2020) results were obtained using the FastSurferVINN algorithm from the official GitHub repository (https://github.com/Deep-MI/FastSurfer) in the Google Colab version. For the SynthSeg (Billot et al., 2023a; Billot et al., 2023b) results, we utilized the code available at https://github.com/BBillot/SynthSeg, employing the robust 2.0 model. The entire segmentation process was conducted on the Google Colab platform. The segmentation results were visualized using four slices that covered the region from the bottom to the top of the basal ganglia. This section encompassed most of the critical structures of interest, allowing comprehensive examination and evaluation.

The public *blind* evaluation took place on May 13, 2023, via a public Twitter post (https://twitter.com/FangChengYeh/status/1657426341803827201) and remained open for 4 days. The evaluators were tasked with providing rankings for the three methods under evaluation. Each response was meticulously coded as three paired comparisons. For instance, if an evaluator expressed a preference for Method A over Method B and Method C, the response was coded as A>B, A>C, and B>C. In cases where an evaluator identified only the best method without ranking the remaining two, we assigned equal ranks to the omitted methods. When two methods were deemed of equal quality, a weight of 0.5 was assigned to each comparison (e.g., A=B was coded as 0.5 for A>B and 0.5 for B>A). Furthermore, some evaluators submitted two responses under different evaluation conditions. In such cases, each response was assigned a weight of 0.5 to ensure equal weighting during analysis. By adding up the total number of responses for each of the six possible pairwise comparisons (A>B, A>C, B>A, B>C, C>A, C>B), we obtained insights into the preferences of the evaluators.

At the end of the evaluation, we received 36 responses. Each evaluator was identified by searching their ID using the Google search engine to confirm their academic positions and training backgrounds. One response was excluded as the individual could not be identified. We successfully identified the remaining 35 evaluators. They all had backgrounds in neuroscience, brain imaging, and image processing. Notably, 29 evaluators held advanced degrees, such as MD or PhD.

***Code and data availability***

The augmentation steps were implemented using the C++ programming language and the Template Image Processing Library (TIPL) to utilize multi-core CPU computation. The source code for all the visual perception augmentation steps is available at: https://github.com/frankyeh/UNet-Studio/blob/main/visual_perception_augmentation.cpp. The visual perception augmentation and the 3D U-Net are integrated into U-Net Studio (http://unet-studio.labsolver.org/) with source code available at https://github.com/frankyeh/UNet-Studio. The repository documented the development process of the code over time to allow for transparent research. The training data were also available at https://github.com/frankyeh/UNet-Studio-Data. The models for each species can be reproduced using the data and tool provided above.

# Results

### *Visual perception augmentation*

Fig. 1 shows the overview of template-based training and exemplary results of visual perception augmentation using a rhesus monkey template as an example. In Fig. 1a, the process starts with the augmentation of a single template image, followed by training a regular 3D U-Net model, estimating the loss, and performing backpropagation (Online Methods).

Fig. 1b provides a more detailed view of the visual perception augmentation steps, whereas Fig. 1c presents exemplary results of each augmentation step. The process begins with image reduction, where the template image undergoes image subsampling and the addition of uniform noise. Additionally, an optional cropping step extracts a circular region from the image and label. The cropping step assumes that the target can be recognized by parts, making it suitable for specific conditions where an anomaly or incomplete region is present. It is important to note that subsampling and adding uniform noise are applied only to the image without modifying the label, whereas cropping is simultaneously performed on both the image and label.

Next, the augmented images enter the lighting step, as depicted in Fig. 1b, with exemplary results in Fig. 1c. This step modifies the image's intensity by manipulating ambient, diffuse, and specular light. These lighting effects solely affect the image, leaving the label unchanged. Then, the data are further augmented by rigid motion and camera transforms to simulate diverse viewing directions and expand the model's capacity to recognize objects and structures from different angles and perspectives. Specific limits are imposed on the rotation angle, translocation distance, and other parameters used in camera transform (Online Methods). As mentioned in the Introduction section, these limitations were set to consider *viewport dependence*. We carefully calibrated the augmentation parameters so that the brain region remains recognizable as determined by a neuroanatomist (the first author). After the rigid motion and camera transformation, the background of the template image (determined by the label) is further augmented by artificial textures, including stamping and Perlin textures. These synthetic background textures are exclusively added to the image background to increase diversity and complexity in the training data. This background augmentation does not modify the label.

We further visualize the results from all visual perception augmentation steps applied to the human template (Fig. 2) and the rhesus monkey (Suppl. Fig. 1), respectively. Fig. 2 presents the initial set of 64 augmented images and labels. Each pair in the figure showcases an augmented image on the left and its corresponding segmentation label on the right. As depicted in the figure, the augmentations enhance the foreground of the template image through different illuminations and geometry transformations, while the background is populated with various artificial textures. It is important to emphasize that each augmented image is used only once during training, and the training process generates 168,000 augmented images from the same template.

***Training and evaluation errors***

We further quantitatively examined the impact of each visual perception augmentation step on evaluation errors. Since only one template image is used in training, we cannot apply cross-validation to estimate the testing error. Therefore, we employed two surrogate approaches to evaluate the model performance: the first is the *template error* generated by evaluating the model using the original template image before augmentation. The template error can be viewed as a converged version of the training error since it is calculated using the same unaugmented template image at every epoch. The second surrogate is the *evaluation error* calculated from a subject's unseen evaluation scan image. Its segmentation label is often unavailable and can be generated by another independent source. This evaluation error is similar to the test error because the evaluation image is not used in training or augmentation; however, the evaluation error will be slightly higher than the test error because of the discrepancies between labeling sources (the template labels and the segmentation labels here are generated from different sources and may differ due to observer differences).

Fig. 3a shows an example axial view of the template used in training highlighted within the orange-brown box, whereas an example view of the evaluation image from the Human Connectome Project (ID #100206) is shown by the blue box. The segmentation of the evaluation image was obtained from SynthSeg(Billot et al., 2023a), a brain segmentation tool based on 3D U-Net. In all inset figures of Fig. 3, the errors are reported by mean squared error (MSE) at the logarithmic scale. The gray line depicts the training error derived from the U-Net output. Additionally, the errors derived from the unaugmented template image (referred to as the *template error* henceforth) were further delineated by two lines: one calculated from foreground voxels (represented by the orange line) and another calculated from background voxels (represented by the brown line). On the other hand, the evaluation errors are illustrated by the blue and dark blue lines, representing foreground and background voxels, respectively.

Fig. 3b shows the training and evaluation errors when the rigid motions were used as only the augmentation approach. The 3D U-Net was trained from scratch with 1,000 epochs and a learning rate of 0.001. Each epoch used only one augmented template. Although the training errors show a continuously decreasing trend, overfitting happens early, as revealed by a large gap between the template error and evaluation errors in both the foreground and background regions. This suggests that a simple translocation and rotation have a very limited data augmentation effect, and the performance is not ideal due to overfitting.

Fig. 3c demonstrates the errors using rigid motion and camera transformations as augmentation approaches. The settings used in Fig. 3c were identical to those in Fig. 3b, except for the additional camera transformation step encompassing image scaling, aspect ratio adjustment, perspective transformation, and lens distortion. Comparing Fig. 3c to Fig. 3b, the error pattern remains essentially unchanged, and notable gaps persist between the template and evaluation errors. Despite the potential for rigid motion and camera transformations to generate a vast array of substantially different images, our findings underscore their limited effectiveness as standalone methods for data augmentation.

Fig. 3d illustrates the errors using rigid motion and image reduction. The settings employed in Fig. 3d were the same as those in Fig. 3b, except for the image reduction process. The image reduction here involved subsampling and the addition of uniform noise applied with a 50% chance. Comparing Fig. 3d to Fig. 3b, the application of image reduction demonstrates visible improvement in the evaluation errors in the background. Yet, the foreground errors remain unchanged, and there are still significant gaps between the template errors and evaluation errors.

Fig. 3e shows the errors using rigid motion and image cropping. The settings in Fig. 3e were the same as those in Fig. 3b, except for the cropping process, which was applied randomly at a 50% chance. It is important to note that cropping modifies the image labels, unlike other image reductions such as subsampling or uniform noise, as observed in Fig. 3d. Comparing Fig. 3e to Fig. 3b, image cropping, similar to other image reduction steps, leads to an improvement in background errors when dealing with unseen subject data. This improvement seems more pronounced than the other image reductions shown in Fig. 3d. However, there is no noticeable improvement in the foreground error. The substantial gap between the template and evaluation errors remain unchanged.

Fig. 3f illustrates the errors using rigid motion and lighting effects. The settings in Fig. 3e were the same as those in Fig. 3b, except for the lighting effects. The lighting effects consisted of ambient, diffuse, and specular light; each was applied randomly and independently with a 50% chance. The introduction of lighting effects in Fig. 3f substantially improves the performance in dealing with unseen data, as shown by the reduced evaluation errors. The foreground errors of the template and evaluation images largely overlap. The background evaluation errors also present a substantial reduction. These results suggest that lighting has a profound impact as a data augmentation technique and substantially enhances the trained network's ability to handle the foreground region of the unseen images.

Fig. 3g depicts the errors using rigid motion and adding background texture through stamping or Perlin texture. The settings in Fig. 3g were the same as those in Fig. 3b, except for the background texture applied randomly at a 50% chance. While introducing background texture in Fig. 3g improves the handling of image backgrounds, it is essential to note that overfitting becomes evident at approximately the 400th epoch, as shown by the increasing gap between the template and evaluation errors.

Fig. 3h and 3i present the errors using rigid motion, camera transformation, image reduction, lighting, and background textures. Fig. 3i further incorporates cropping to consider incomplete foreground. Both figures demonstrate significant improvements in the foreground and background errors: the errors derived from the template and evaluation images largely overlap, indicating that visual perception augmentation effectively mitigates the overfitting issue. The template errors closely align with the evaluation errors observed with unseen data. This result suggests that all augmentation approaches synergistically enhance the trained network's adaptability in handling unseen data.

Next, we applied full-scale training to construct working models for brain tissue segmentation. Suppl. Fig. 2 presents the training and template errors observed during three training rounds for the human, rhesus monkey, marmoset, rat, and mouse models. Each model was trained using the specific template corresponding to its respective species, facilitating species-specific learning. All visual perception augmentation steps except for cropping were applied. In the first round, the model was trained for 1000 epochs, with 8 augmented images accumulated per epoch, employing a learning rate of 0.001. The second and third rounds comprised 2000 epochs each, with 16 and 32 augmented images accumulated per epoch, respectively. The learning rates for these rounds were set to 0.005 and 0.0025, respectively. The overall training and template errors consistently decrease across all species, indicating that this template-based training approach can be applied to different templates, and the training models are progressively converging over time.

### *Transfer training*

Due to unable to find a suitable mouse template and label, we adopted a *transfer training* approach to train our mouse model. Suppl. Fig. 3 presents an overview of this transfer training approach that trains the mouse model by the label output from the *rat* model. We first utilized a set of 11 publicly available mouse scans (Suppl. Fig. 3a) and performed an averaging process to create a representative population-averaged template tailored to mice (depicted in Suppl. Fig. 3b). To segment the mouse template, we employed the already trained rat model (Suppl. Fig. 3c). Remarkably, the rat model, in its vanilla form without transfer learning, required no further modifications to segment the mouse template effectively. By utilizing the mouse template and the segmentation derived from the rat model, we applied the same set of visual perception augmentation techniques to train the U-Net model (Suppl. Fig. 3d). Our template-based transfer training approach highlights its adaptability in facilitating knowledge transfer and effective modeling across species.

### *Evaluation on human, rhesus monkey, marmoset, rat, and mouse data*

Fig. 4 presents the evaluation results of our models combined with post-processing on different species. The human, rhesus monkey, marmoset, and rat models were trained using each of their dedicated templates and corresponding labels (Suppl. Table 1). The output was further utilized by the post-processing routine (Online Methods) to generate (1) skull-stripped images, which removes signals outside the brain, (2) segmentation labels, which offers voxel-wise label of the five tissue types, and (3) tissue maps of white matter and gray matter, which provide volumes of the tissue probability. The specific outputs for white matter and gray matter segmentations can be observed in the last two rows of Fig 4, whereas more detailed tissue segmentation results on the human T1W image are illustrated in Fig. 5.

We evaluated the models using individual scan data not utilized during training. The evaluation images were selected from the first scans of each dataset (Suppl. Table 2). As shown in Fig. 4 and Fig. 5, the overall result demonstrates the effectiveness of template-based training in delineating different brain structures, even with the limited training data consisting of a single template image. Remarkably, these models achieved accurate segmentations despite being trained solely on a single template image.

### *Comparison with other U-Net methods*

We further compared the performance of the proposed method with FastSurfer(Henschel et al., 2020) and SynthSeg(Billot et al., 2023b) on the first young adult subject (#100206) from the Human Connectome Project in Fig 6. All these approaches leveraged a U-Net architecture. To facilitate comparison, we merged FastSurfer and SynSeg segmentation results into the same tissue segmentation label used in this study, including white matter, gray matter, cerebellar cortex, basal ganglia, and others. Fig. 6 shows four axial slices surrounding the basal ganglion, automatically chosen to maintain an equal gap of 16 slices. These slices are from FastSurfer (Fig. 6a), SynthSeg (Fig. 6b), and the proposed method (Fig. 6c). There are no ground truth labels in this human scan, and many imaging voxels may contain more than one tissue type due to the partial volume effect. Thus, we resorted to expert evaluations in this comparison, and a public blind evaluation was organized (Online Methods).

The evaluation results are listed in Suppl. Table 3. A total of 35 identifiable evaluators joined the effort. All of the evaluators have experience in brain imaging studies, and 29 of them have advanced degrees. The voting results showed that 12, 2, and 21 evaluators recommended A (FastSurfer), B (SynthSeg), and C (Proposed method) as the best method. A more detailed pairwise comparison shows that the comparison between A and B received 24.5 and 8.5 votes, respectively. The comparison between A and C received 9.5 and 23.5 votes, respectively. The comparison between B and C received 5 and 27 votes, respectively. Although public blind evaluation results only reflect the public interests, not necessarily the ground truth, the fact that the template-based training (method C) received the most favorable results suggests that its performance is not visually inferior to the existing brain segmentation tools.

### *Evaluation using animal data*

Additionally, we evaluated the model performance on various acquisition conditions to gauge the limitations. Animal images, unlike the standardized clinical scans of human subjects, are known to have very diverse acquisition settings. Image acquisition for animal studies often encompasses variations in pulse sequences, imaging resolution, receiver coils, and post-acquisition reconstruction techniques. Here we tested our models on various animal acquisitions to break the trained models' limits and understand the conditions they fail.

We began evaluating our rhesus monkey model on all 20 data sources participating in the PRIMatE Data Exchange (PRIME-DE) consortium(Milham et al., 2018). Due to the numerous scans available, we selected only the first scan data from each data acquisition site to assess the robustness of our trained model. Suppl. Fig. 5 shows the segmentation results on rhesus monkey data in mid-sagittal slices. We present results at mid-sagittal slices due to their ability to provide comprehensive observations of acquisition diversity, encompassing factors such as head position, coil sensitivity, signal inhomogeneity, signal contrast, and image resolution. The results demonstrate grossly correct segmentation, even under non-isotropic resolution and prominent inhomogeneity. However, two notable errors can be identified. An error was observed below the cerebellum at the $2^{nd}$ column of the $1^{st}$ row, likely due to a significant posterior rotation. This error could be mitigated by rotating the image volume to match the training template. Another error was noticed at the $2^{nd}$ column of the $2^{nd}$ row, where the segmentation of the frontal region included high-intensity tissue outside the brain. This error may be attributed to significant signal inhomogeneity, as the input images were not processed with inhomogeneity correction.

Furthermore, we evaluated our rat model using the recent standard rat T2-weighted 2D-slice data acquired by 20 research groups(Grandjean et al., 2023) and shared on the open neuro website (https://openneuro.org/datasets/ds004116)(Markiewicz et al., 2021). Suppl. Fig. 6 showcases the segmentation results on rat data at mid-coronal slices, as the T2-weighted images were acquired using 2D coronal slices. The overall results demonstrate correct segmentation, regardless of different inhomogeneities, and no obvious errors were observed, likely due to a standardized protocol. While the trained model exhibits a certain level of robustness,

Last, we evaluated our mouse model trained using all available data shared on the open neuro website (https://openneuro.org/)(Markiewicz et al., 2021). All images from 5 datasets were used to evaluate the model. The images have diverse acquisition settings, including the contrast pattern, resolution, and positions, as shown in Suppl. Fig. 7.  In the 1st dataset, segmentation errors can be observed at locations with substantial distortion and artifacts. There is no obvious segmentation error found in the 2$^{nd}$ and 4$^{th}$ datasets. In the 3$^{rd}$ dataset, additional background segmentation was found in the last scan, likely due to substantial left-right rotation. In the 5$^{th}$ dataset, the segmentations are coarse and not accurate enough, probably due to the substantially low resolution of the images.

Overall, we identified limitations of our models: the prominent noise and artifact, substantial rotation, and large reduced resolution may cause segmentation errors. These findings pointed to further refinement and improvements in the segmentation process, particularly in the context of animal imaging data. Additional preprocessing and quality-control steps can be added to mitigate the challenges.

***Cross-species evaluation***

Last we examined the generalization limits of the models. The models evaluated here were trained by one template and did not receive further improvement using transfer learning. Fig. 7a provides insights into the generalization limits of the human model applied to chimpanzee (1$^{st}$ row) and rhesus monkey MRI (2$^{nd}$ row). The sources of these two evaluation images are listed in Suppl. Table 2. Both test images were selected from the first scan data of the example dataset. As shown in the figure, the skull stripping in the chimpanzee image displays no obvious errors, and the result on the rhesus monkey is grossly correct, except for imperfect segmentation at the frontal tip. On the other hand, the segmentation performs grossly well in the separation of gray and white matter, except for an error in labeling the putamen as gray matter in non-human primates.

Fig. 7b examines the generalization limits of the rhesus monkey model applied to human, chimpanzee, and marmoset data. The rhesus monkey model here was only trained by the rhesus template and did not receive improvement using transfer learning. As shown in the figure, the rhesus model shows satisfactory results in skull stripping on all species. On the other hand, the tissue segmentation, white matter map, and gray matter map show grossly correct segmentation except for the error in the human putamen. The overall results suggest the rhesus monkey model, without using transfer learning, already exhibits good generalization capability in non-human primates.

***Evaluation using brain tumor images***

In Fig. 8, we examine a human model trained to handle incomplete foregrounds. This model was trained exclusively on the ICBM-152 T1W image, and no brain tumor image was used in training. The augmentation steps included F to cope with the incomplete foreground. We selected the first five cases (#2, #6, #8, #9, #11) from the UPenn GBM dataset(Pollard et al., 2020) to evaluate the performance. The T1W images of these cases are shown in the first column, whereas the segmentation results obtained using our trained human model are represented in the second and third columns. The white, blue, and orange contours indicate the brain contour, gray-white matter junction, and basal ganglia. On the other hand, the expert segmentation provided by the original dataset is shown in the last column and denoted by white and blue contours, representing the tumor region and peritumoral edema, respectively. As shown in the figure, the segmentation results on #2, #6, #9, and #11 present a black empty region corresponding to the tumor locations identified by expert labels. The model isolates these anomalies regions while segmenting the remaining brain tissues. The tumor size of case #8 is smaller than other cases, and the trained model failed to isolate the lesion. There are also substantial differences between the anomaly detected by the models and those segmented by experts. Overall, the model shows capabilities in separating large lesions, despite being trained solely on a template without using any brain tumor images.

## Discussion

In this study, we trained U-Net models to provide consistent skull-stripping and tissue segmentation in multiple species, including mouse, rat, marmoset, rhesus, and human brains. Utilizing a novel augmentation approach, we trained these models using only a single template image, overcoming the challenge of limited training data availability. Extensive evaluations allowed us to gain insights into the strengths and limitations of the template-based training approach. The trained models offer a unified solution for researchers to segment brain tissues and investigate brain structures across species in imaging neuroscience. Moreover, the template-based training approach represents a significant contribution, as it effectively addresses data availability challenges and opens new opportunities for research and discovery that were previously unfeasible due to limited amount of training data.

The primary contribution of this study is a set of U-Net models that consistently perform brain tissue segmentation across commonly studied species in neuroscience. To achieve this, we employed a novel augmentation technique, as described earlier, which enabled our models to adapt to diverse visual input conditions and address challenges associated with viewpoint dependence. By incorporating a comprehensive range of image augmentation methods, such as reduction, noise introduction, lighting simulation, geometry and camera transformations, and background textures, our models were effectively equipped to handle varying imaging settings and occlusions. As a result, they demonstrated robustness and accuracy in segmenting brain tissues, thereby allowing researchers to analyze and compare brain structures across different species in a reliable manner.

The second significant contribution of this study lies in the methodology of training a deep neural network using just one template image. Traditional deep learning approaches for image segmentation heavily rely on abundant labeled datasets(Keshari et al., 2020; LeCun et al., 2015) and often employ techniques like image rotation and flipping to augment the training data(Yang et al., 2022). However, these augmentation methods are insufficient, as shown in our Result section. In contrast, the comprehensive augmentation strategies developed in this study demonstrated the feasibility of training a 3D U-Net model using a single template image. This allowed future studies to construct new models for various animal species and to overcome the challenges posed by limited data in multi-species biomedical research.

Another noteworthy finding in this study is the remarkable generalization ability demonstrated by the U-Net models trained using visual perception augmentation. Specifically, without transfer learning, the rhesus monkey model effectively generalized to other non-human primate data and produced satisfactory results. Similarly, the rat model trained in this study successfully provided segmentation labels for mouse data, enabling us to train the mouse model. This generalization capability alleviates the burden of manual labeling by domain experts and ensures consistent segmentation outcomes across different species. By establishing a consistent segmentation framework, we can identify shared mechanisms and patterns that transcend species boundaries, thereby enhancing our understanding of the brain and its intricate functions. This finding has significant implications for advancing comparative neuroscience and facilitating cross-species analysis, ultimately contributing to the broader comprehension of brain structure and function across different species.

The potential of template-based training in anomaly detection is an area worth exploring further. Although we are yet to explore its full potential, in our study, we have demonstrated that a U-Net model, trained solely on one template image, can isolate anomalies without the need for lesion-specific training data. The ability to detect lesions opens exciting possibilities for efficiently screening large datasets without using lesion-specific examples during training. This breakthrough has promising applications in lesion detection, quantification, and treatment planning, as it streamlines the identification process and has broader implications for improving various aspects of healthcare. By reducing the reliance on a large number of lesion-specific training examples, this approach can potentially enhance medical image analysis, ultimately leading to better patient care.

We have integrated the outcome of this study into a comprehensive and user-friendly tool known as U-Net Studio (https://unet-studio.labsolver.org). U-Net Studio is an inclusive and user-friendly platform that seamlessly integrates the template-based training framework and all trained models developed in this study. It provides researchers with a versatile interface that not only facilitates segmentation tasks on commonly studied animal species but also empowers them to construct new models for addressing novel applications in neuroscience research. By making U-Net Studio an open-source tool, we further aim to foster collaboration and inspire innovation within the scientific community, thereby promoting significant advancements in the field of neuroscience research.

It is essential to acknowledge the inherent limitations of our methods. In addition to the limitations presented in the Results section, one critical limitation is the reliance on accurate labels for training the models. The models can learn and reproduce flawed segmentations if the template labels contain errors or inaccuracies. Moreover, obtaining high-quality template images with reliable and precise labels is challenging. In our study, we encountered difficulties in finding a suitable mouse template with satisfactory segmentation results. The available mouse templates either lacked a background or associated tissue segmentation labels. When using a problematic template or label data, the training would take much more time to converge and template errors tend to fluctuate instead of presenting a decreasing pattern. These are signs indicating issues in the training data that need expert inspection and corrections. In this study, we identified issues with publicly available templates or labels, necessitating manual corrections (e.g., modifications listed in Suppl. Table 1). These limitations underscore the importance of careful selection and meticulous preparation of template data to ensure the accurate and reliable training of deep learning models.

Looking ahead, there are opportunities to enhance our methodology by incorporating additional preprocessing routines to address challenges related to rotation and signal inhomogeneity. Moreover, integrating other aspects of human visual perception, such as depth perception and object recognition under occlusion, can further improve the models' ability to learn and generalize across different species. Future studies can also explore the integration of transfer learning techniques with visual perception augmentation to enhance the models' generalization capabilities and reduce reliance on a single template image. By leveraging transfer learning, the models can adapt to new datasets and improve performance on previously unseen data, bolstering their versatility and applicability in various research and practical applications. These advancements hold great potential to advance the field of brain tissue segmentation and open new avenues for research and discovery in neuroscience and related disciplines.

# References

Ahmad, S., Wu, Y., Wu, Z., Thung, K.H., Liu, S., Lin, W., Li, G., Wang, L., Yap, P.T., 2023. Multifaceted atlases of the human brain in its infancy. Nat Methods 20, 55-64.

Billot, B., Greve, D.N., Puonti, O., Thielscher, A., Van Leemput, K., Fischl, B., Dalca, A.V., Iglesias, J.E., 2023a. SynthSeg: Segmentation of brain MRI scans of any contrast and resolution without retraining. Medical image analysis 86, 102789.

Billot, B., Magdamo, C., Cheng, Y., Arnold, S.E., Das, S., Iglesias, J.E., 2023b. Robust machine learning segmentation for large-scale analysis of heterogeneous clinical brain MRI datasets. Proceedings of the National Academy of Sciences 120, e2216399120.

Ciric, R., Thompson, W.H., Lorenz, R., Goncalves, M., MacNicol, E.E., Markiewicz, C.J., Halchenko, Y.O., Ghosh, S.S., Gorgolewski, K.J., Poldrack, R.A., Esteban, O., 2022. TemplateFlow: FAIR-sharing of multi-scale, multi-species brain models. Nat Methods 19, 1568-1571.

DeVries, T., Taylor, G.W., 2017. Improved regularization of convolutional neural networks with cutout. arXiv preprint arXiv:1708.04552.
Grandjean, J., Desrosiers-Gregoire, G., Anckaerts, C., Angeles-Valdez, D., Ayad, F., Barrière, D.A., Blockx, I., Bortel, A., Broadwater, M., Cardoso, B.M., 2023. A consensus protocol for functional connectivity analysis in the rat brain. Nature neuroscience, 1-9.
Henschel, L., Conjeti, S., Estrada, S., Diers, K., Fischl, B., Reuter, M., 2020. Fastsurfer-a fast and accurate deep learning based neuroimaging pipeline. NeuroImage 219, 117012.
Keshari, R., Ghosh, S., Chhabra, S., Vatsa, M., Singh, R., 2020. Unravelling small sample size problems in the deep learning world, 2020 IEEE Sixth International Conference on Multimedia Big Data (BigMM). IEEE, pp. 134-143.
Krizhevsky, A., Sutskever, I., Hinton, G.E., 2017. Imagenet classification with deep convolutional neural networks. Communications of the ACM 60, 84-90.
LeCun, Y., Bengio, Y., Hinton, G., 2015. Deep learning. nature 521, 436-444.
Markiewicz, C.J., Gorgolewski, K.J., Feingold, F., Blair, R., Halchenko, Y.O., Miller, E., Hardcastle, N., Wexler, J., Esteban, O., Goncavles, M., 2021. The OpenNeuro resource for sharing of neuroscience data. Elife 10, e71774.
Milham, M.P., Ai, L., Koo, B., Xu, T., Amiez, C., Balezeau, F., Baxter, M.G., Blezer, E.L., Brochier, T., Chen, A., 2018. An open resource for non-human primate imaging. Neuron 100, 61-74. e62.
Perlin, K., 1985. An image synthesizer. ACM Siggraph Computer Graphics 19, 287-296.
Pollard, K., Banerjee, J., Doan, X., Wang, J., Guo, X., Allaway, R., Langmead, S., Slobogean, B., Meyer, C.F., Loeb, D.M., 2020. A clinically and genomically annotated nerve sheath tumor biospecimen repository. Scientific Data 7, 184.
Ronneberger, O., Fischer, P., Brox, T., 2015. U-net: Convolutional networks for biomedical image segmentation, Medical Image Computing and Computer-Assisted Intervention–MICCAI 2015: 18th International Conference, Munich, Germany, October 5-9, 2015, Proceedings, Part III 18. Springer, pp. 234-241.
Shen, D., Wu, G., Suk, H.-I., 2017. Deep learning in medical image analysis. Annual review of biomedical engineering 19, 221-248.
Tarr, M.J., Pinker, S., 1989. Mental rotation and orientation-dependence in shape recognition. Cognitive psychology 21, 233-282.
Tarr, M.J., Williams, P., Hayward, W.G., Gauthier, I., 1998. Three-dimensional object recognition is viewpoint dependent. Nature neuroscience 1, 275-277.
Thomas, R.M., Bruin, W., Zhutovsky, P., van Wingen, G., 2020. Dealing with missing data, small sample sizes, and heterogeneity in machine learning studies of brain disorders, Machine learning. Elsevier, pp. 249-266.
Thompson, W., Fleming, R., Creem-Regehr, S., Stefanucci, J.K., 2011. Visual perception from a computer graphics perspective. CRC press.
Vass, G., Perlaki, T., 2003. Applying and removing lens distortion in post production, Proceedings of the 2nd Hungarian Conference on Computer Graphics and Geometry, pp. 9-16.
Wagemans, J., Elder, J.H., Kubovy, M., Palmer, S.E., Peterson, M.A., Singh, M., Von der Heydt, R., 2012. A century of Gestalt psychology in visual perception: I. Perceptual grouping and figure–ground organization. Psychological bulletin 138, 1172.
Yang, S., Xiao, W., Zhang, M., Guo, S., Zhao, J., Shen, F., 2022. Image data augmentation for deep learning: A survey. arXiv preprint arXiv:2204.08610.
Zhang, C., Bengio, S., Hardt, M., Recht, B., Vinyals, O., 2021. Understanding deep learning (still) requires rethinking generalization. Communications of the ACM 64, 107-115.

## Acknowledgments

The author was supported by NIH grant R01 NS120954. The MRI data were provided by the Human Connectome Project, WU-Minn Consortium (Principal Investigators: David Van Essen and Kamil Ugurbil; 1U54MH091657), National Chimpanzee Brain Resource (NIH grant NS092988), OpenNeuro (R24MH117179, R24MH114705).

## Author Contributions Statement

FY implemented the code, analyzed data, and wrote the manuscript.

## Competing Interests Statement

The author declares no competing interests.

# Figures

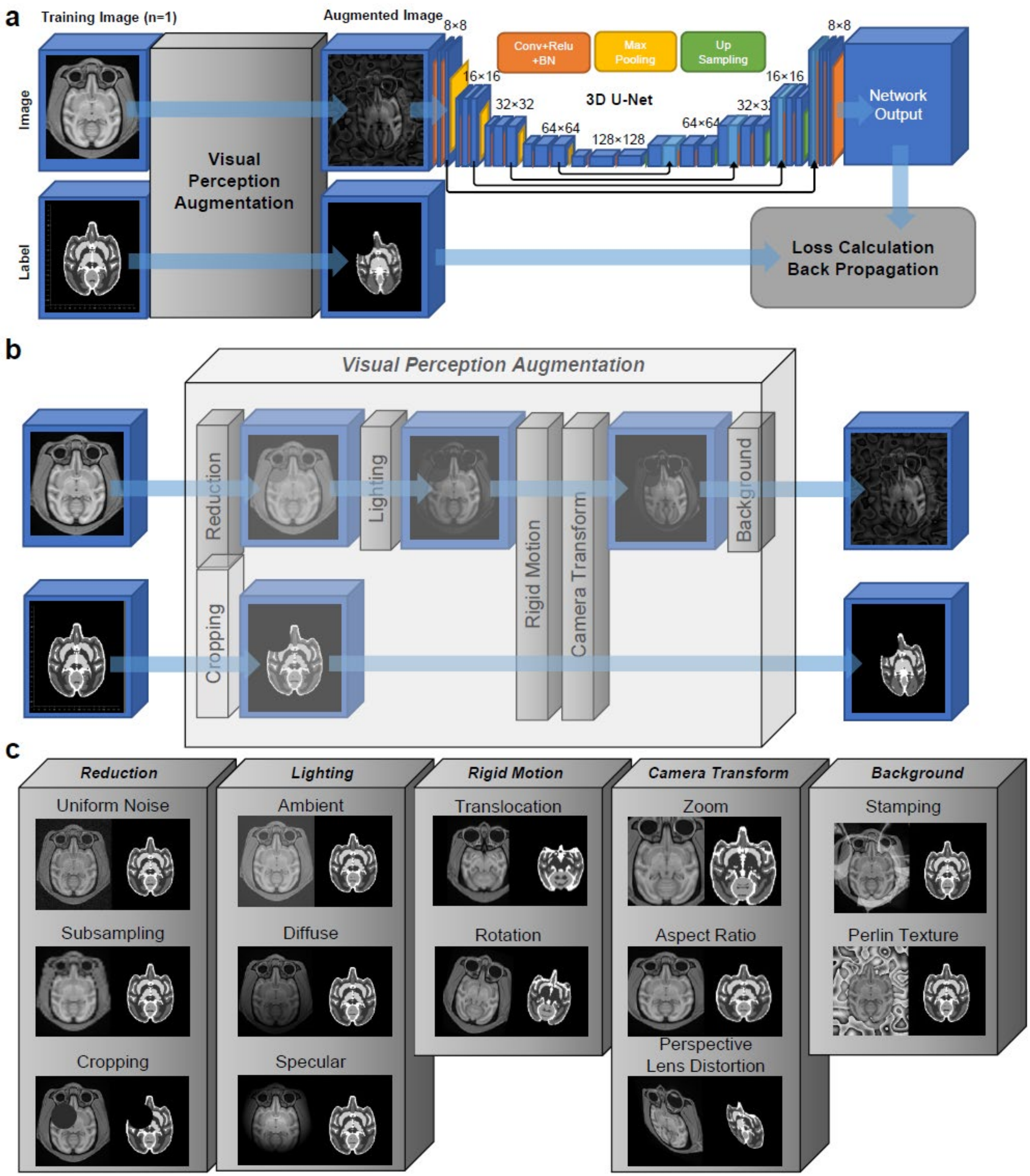

**Figure 1.** Template-based training and exemplary results of visual perception augmentation. (a) The figure illustrates the workflow of the template-based training, which leverages visual

perception augmentation to generate training images for training a 3D U-Net model. Each augmented image is passed through the U-Net, and the output is used for loss estimation and backpropagation. The gradients obtained during backpropagation are accumulated to update the model. (b) The process of visual perception augmentation involves a series of augmentation steps. The first step is image reduction, which incorporates image subsampling, noise addition, and cropping to introduce variability. The lighting step modifies the image intensity to simulate different lighting conditions. The rigid motion and camera transformation steps simulate variations in geometry and viewing directions. Finally, artificial background textures are added to create background variations. (c) The exemplary results of each augmentation step are demonstrated using the rhesus monkey template. These results provide visual examples of the effects of each augmentation step on the template, showcasing the diverse range of images generated through the augmentation process.

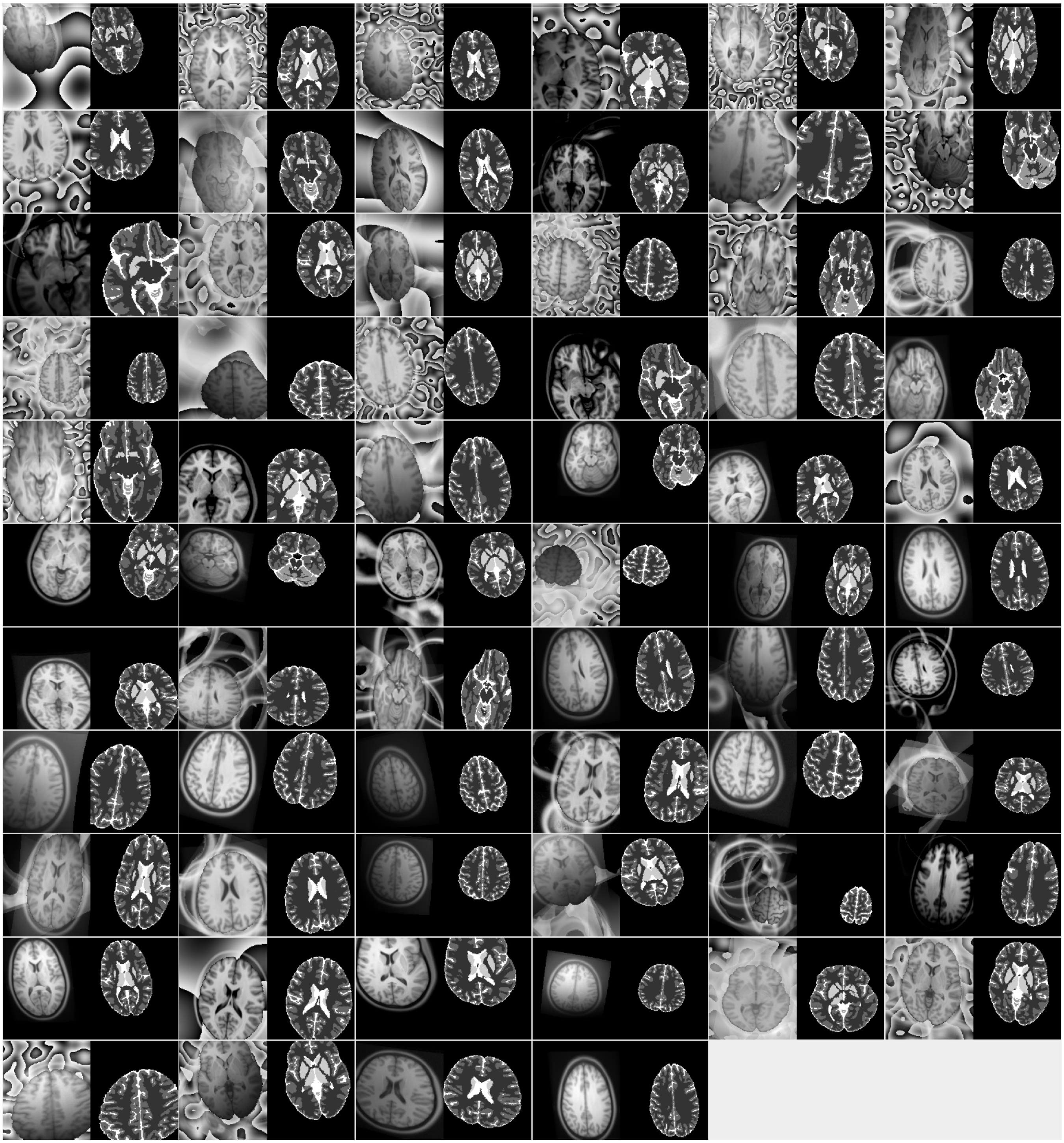

**Figure 2:** The augmented human T1W template using visual perception augmentation. This figure showcases how visual perception augmentation generates the training samples using the same human T1W template. The first 64 image-label pairs resulting from the augmentation process are displayed in the figure. Each pair consists of an augmented image on the left and its corresponding label on the right. The augmentation process encompasses various steps, including image reduction, lighting adjustments, rigid motion simulations, camera transformations, and background texture variations. These augmentations are designed to increase the variability of the training data derived from a single template and enhance the overall robustness of a deep neural network model. It is important to note that each augmented image is utilized only once in training, and a total of 168,000 augmented images will be generated to train the model effectively.

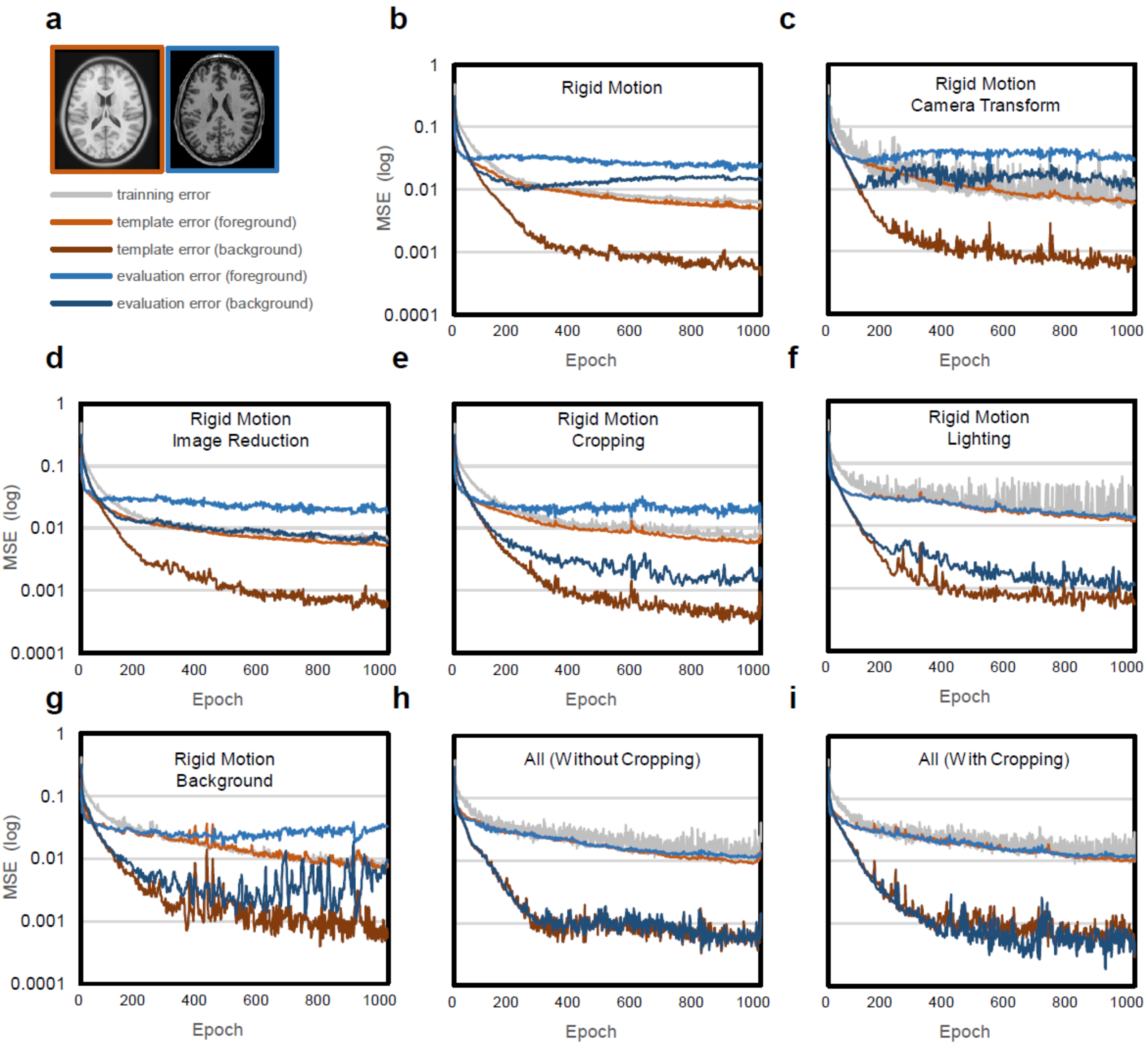

**Figure 3:** Impact of visual perception augmentation on the model performance. (a) The evaluation is conducted using the unaugmented template image (shown in orange) and a subject image (shown in blue) to calculate the template and evaluation errors, respectively. The calculation of errors is further categorized into the background (label=0) and foreground (label$\neq$0) segments. (b) Overfitting is observed when rigid motion is used as the sole augmentation step, as evidenced by the significant gap between the template and evaluation errors. (c) Combining rigid motion and camera transformation steps still has overfitting

problem. (d) Combining rigid motion and image reduction slightly improves evaluation errors in the background but does not have an obvious effect on the foreground. (e) Combining rigid motion and cropping demonstrates improvement in evaluation errors in the background, but there is no notable improvement in the foreground. (f) Combining rigid motion and lighting substantially reduces the overfitting problem, as indicated by the reduced gap between template and evaluation errors. (g) Combining rigid motion and artificial background textures improves the background evaluation errors. (h) Combining all augmentations (excluding cropping) mitigates the overfitting problem, as depicted by the overlap between template and evaluation errors. (i) Combining all augmentations, including cropping, also successfully addresses the overfitting issue. These results demonstrate the impact of visual perception augmentation in reducing evaluation errors and enhancing the model's adaptability to previously unseen data.

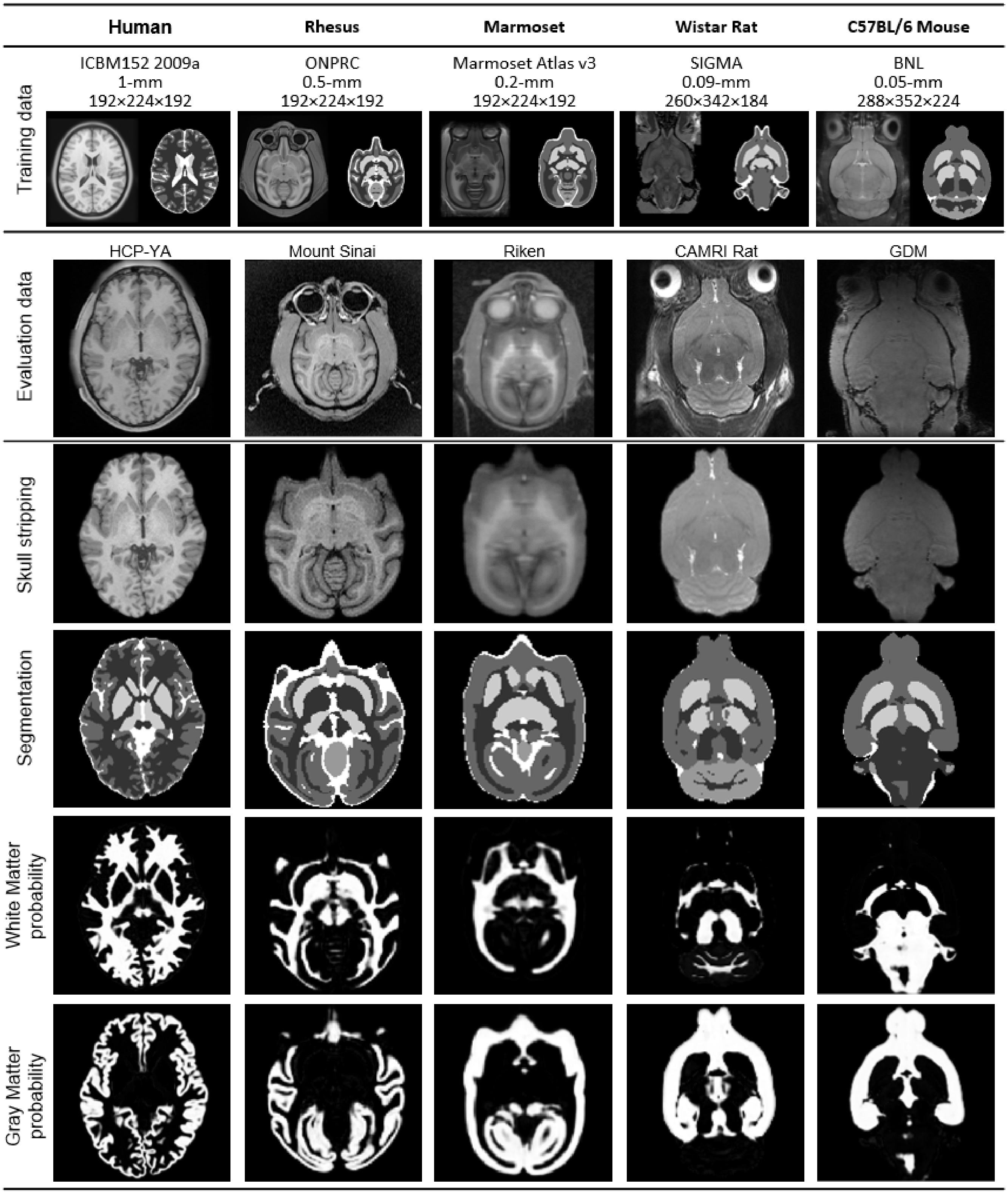

**Figure 4.** Evaluation results showcase the performance on commonly studied species in neuroscience. Each 3D U-Net model was trained using a template and its associated label,

and the model outputs enable skull-stripping, voxel-wise tissue segmentation, and tissue probability map generation. Skull-stripping is achieved by removing the image background through a brain mask generated by summing the U-Net output and applying a threshold of 0.5. Voxel-wise segmentation is obtained by searching the maximum value from the U-Net output. The last two rows of the figure specifically highlight the outputs for white matter and gray matter tissues. These results showcase the effectiveness of the template-based training approach in delineating various brain structures, even when the available data is limited to a single template image.

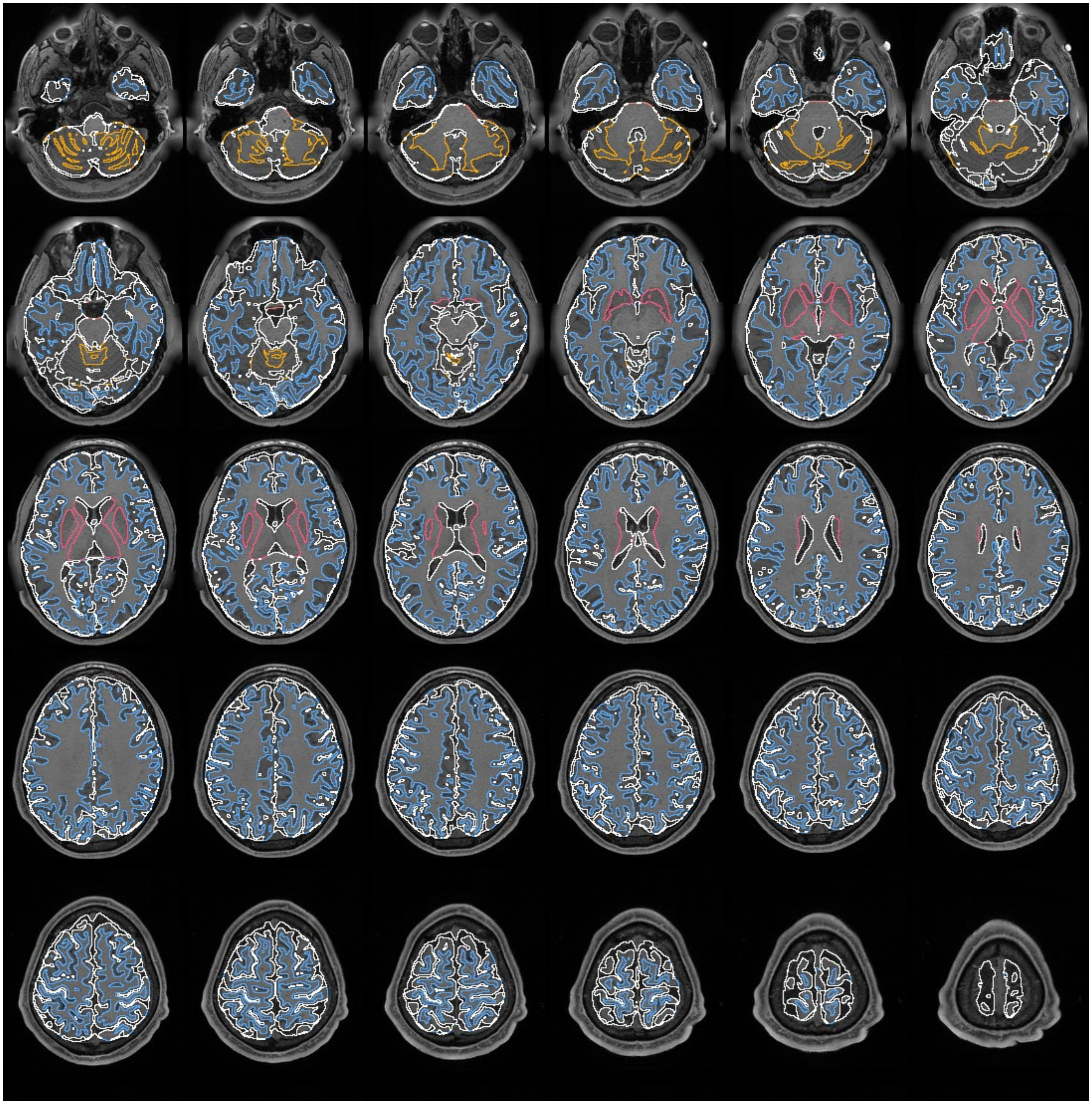

**Figure 5.** Tissue segmentation results on the T1W image of a healthy human young adult using a 3D U-Net trained by the ICBM152 T1W template. The brain surface, a gray-white matter junction, basal ganglia, and cerebellum cortex are delineated by white, blue, red, and yellow lines, respectively. The U-Net model achieves tissue segmentation in a subject's MRI data despite being trained solely on a single population-averaged template image.

a
b
c
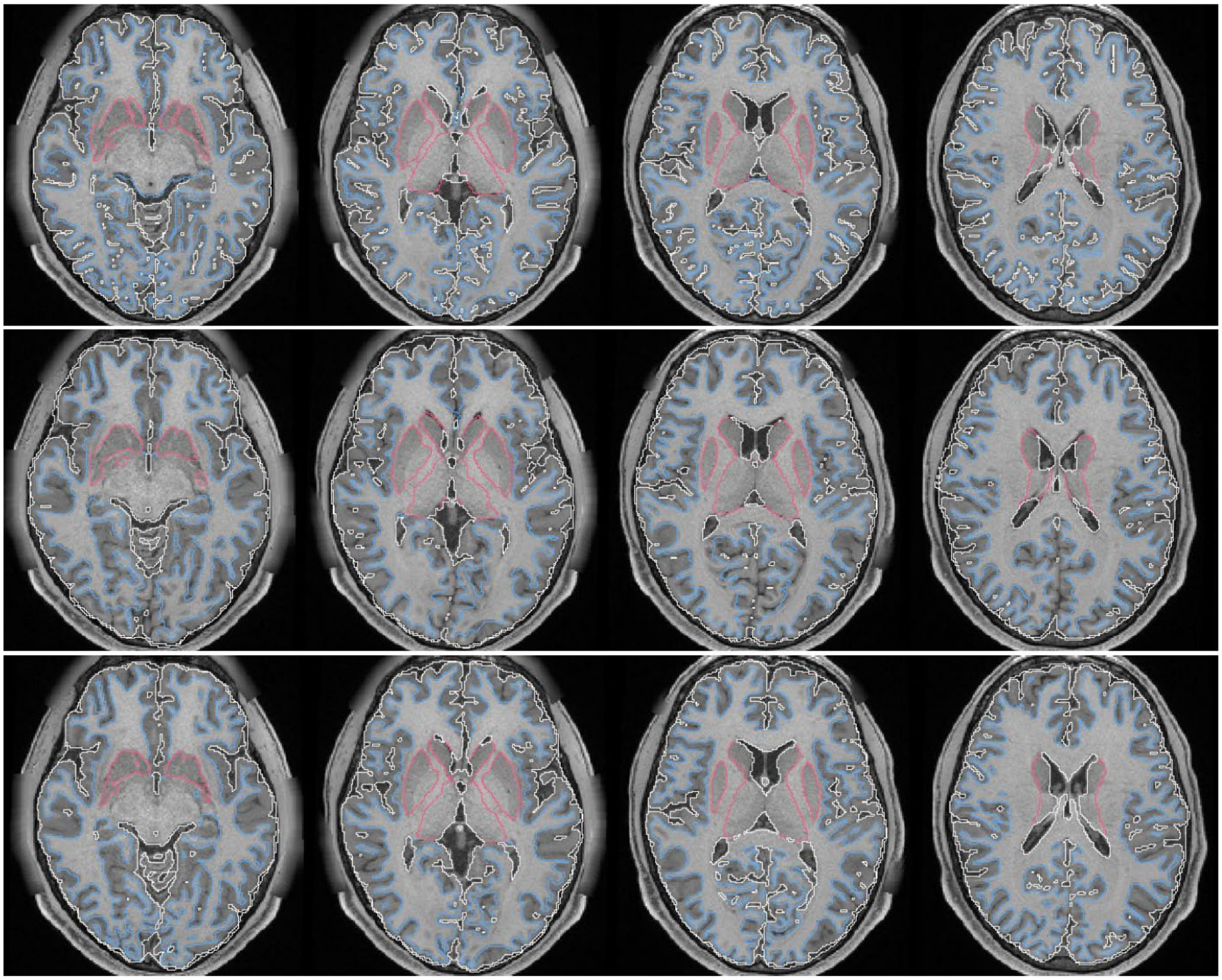

**Figure 6:** Comparison of the different U-Net segmentations using the Human Connectome Project young adult data (ID #100206). The results obtained from three different methods are presented: (a) FastSurfer, (b) SynthSeg, and (c) the model trained in this study. All methods employ a U-Net architecture, and the segmentation output is showcased through five tissue segments: white matter, gray matter, cerebellar cortex, basal ganglia, and others. The white line outlines the surface defined by the gray matter and the brain ventricles. The blue line marks the boundaries between white matter and gray matter. Additionally, the pink line delineates the basal ganglia. A blind assessment was conducted to ascertain expert preference among the various methods.

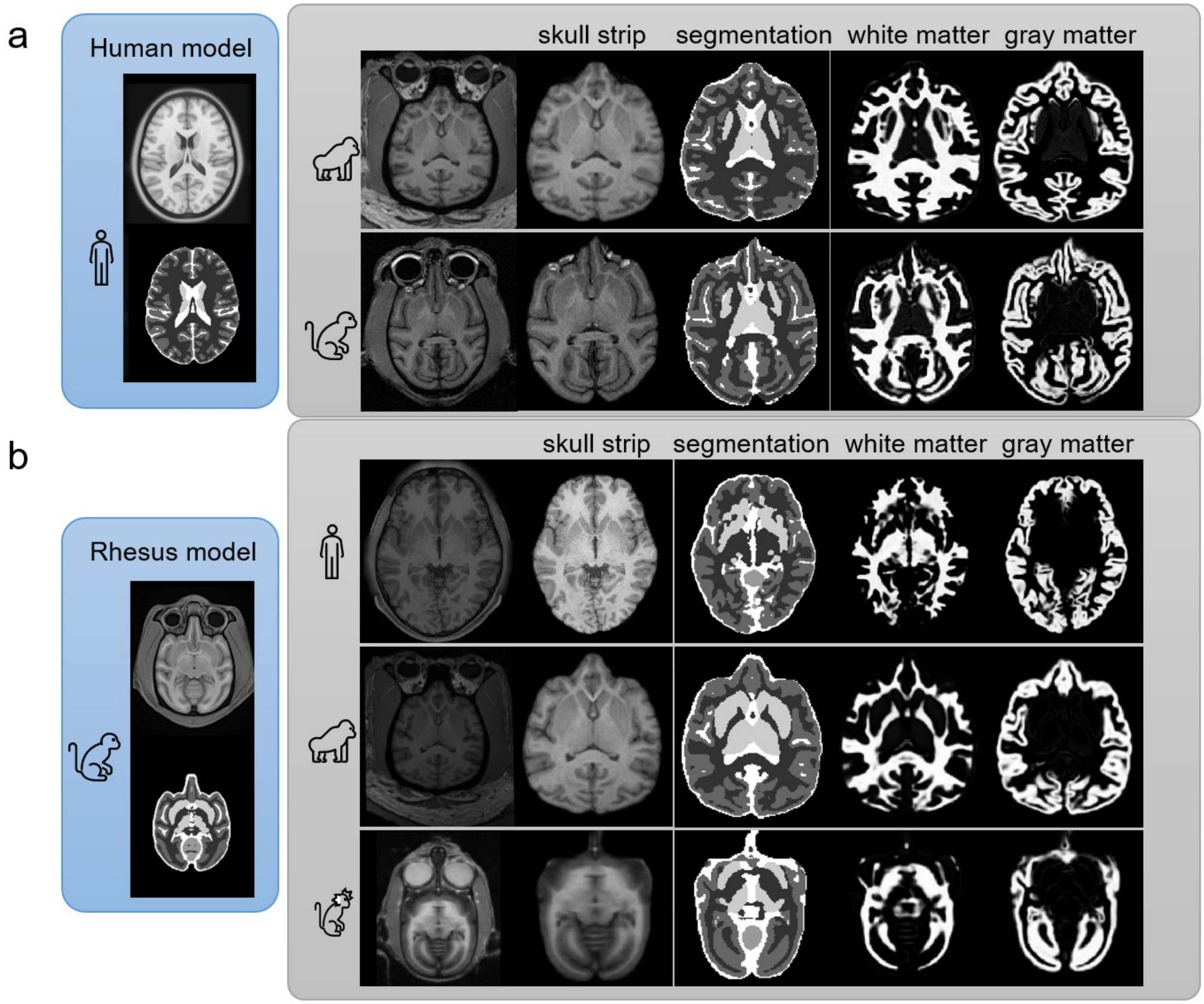

**Figure 7**: Generalization limits of the template-trained models on cross-species segmentation without transfer learning. (a) The 3D U-net model trained using only the human template is applied to chimpanzee and rhesus monkey images. The model successfully accomplishes skull stripping for chimpanzees, while there is a minor over-extended boundary observed in the frontal base of the rhesus monkey. The tissue segmentation can distinguish white matter and gray matter while misclassifying the putamen as gray matter in non-human primates. (b) The 3D U-net model trained using the rhesus monkey template is applied to the T1W image of a human, chimpanzee, and marmoset, respectively. The rhesus monkey model demonstrates no visible error in skull stripping across all three species. The tissue segmentation effectively separates white matter and gray matter in non-human primates.

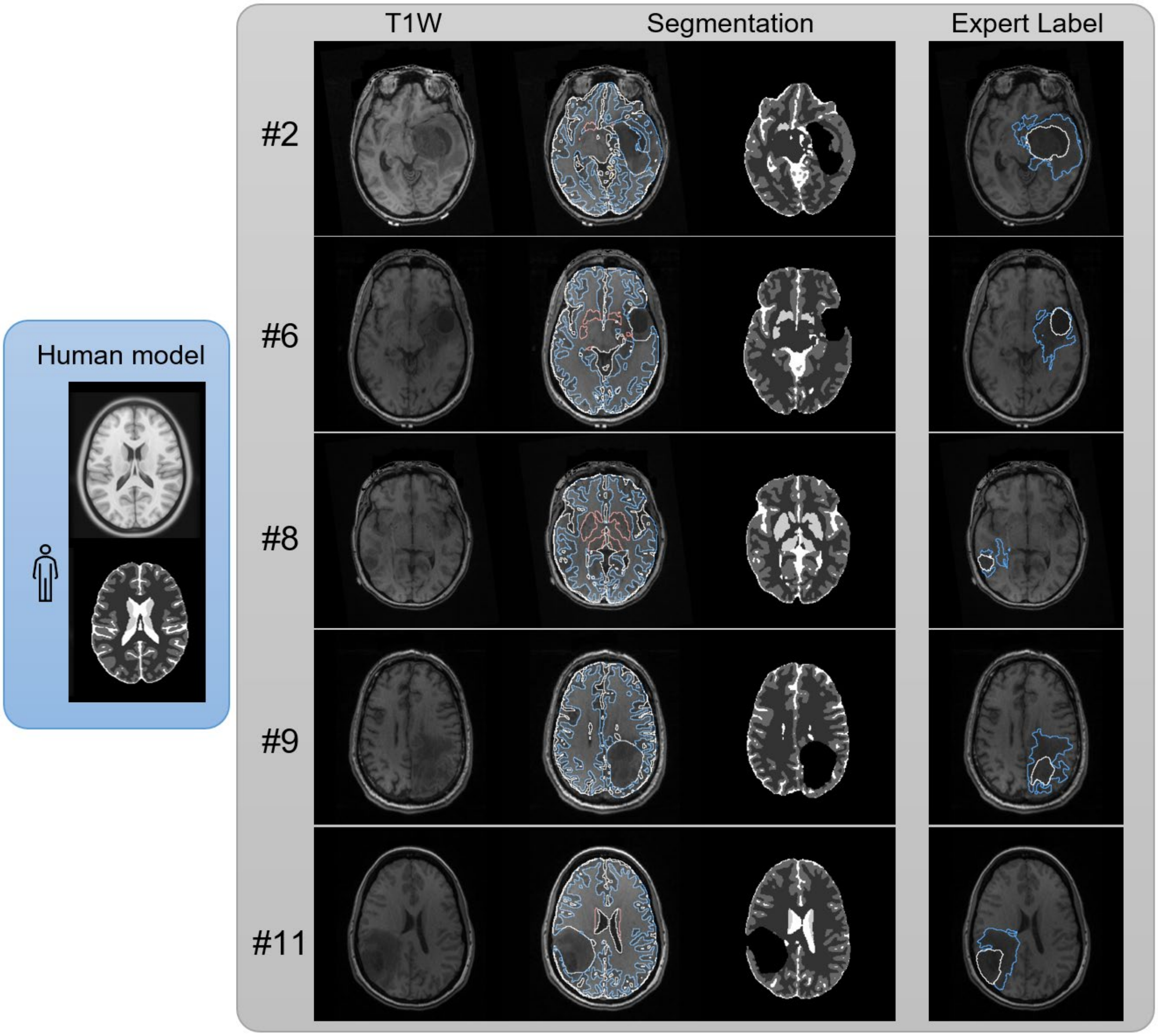

**Figure 8**. Segmentation results on brain tumor images. The model was exclusively trained using the ICBM152 T1W template and did not include tumor or lesion images in its training. The model successfully segments different tissues and presents anomalies in cases #2, #6, #9, and #11 that match the tumor location delineated by experts. The model did not detect the lesion in case #8, likely due to its small size.

# Supplementary Figures

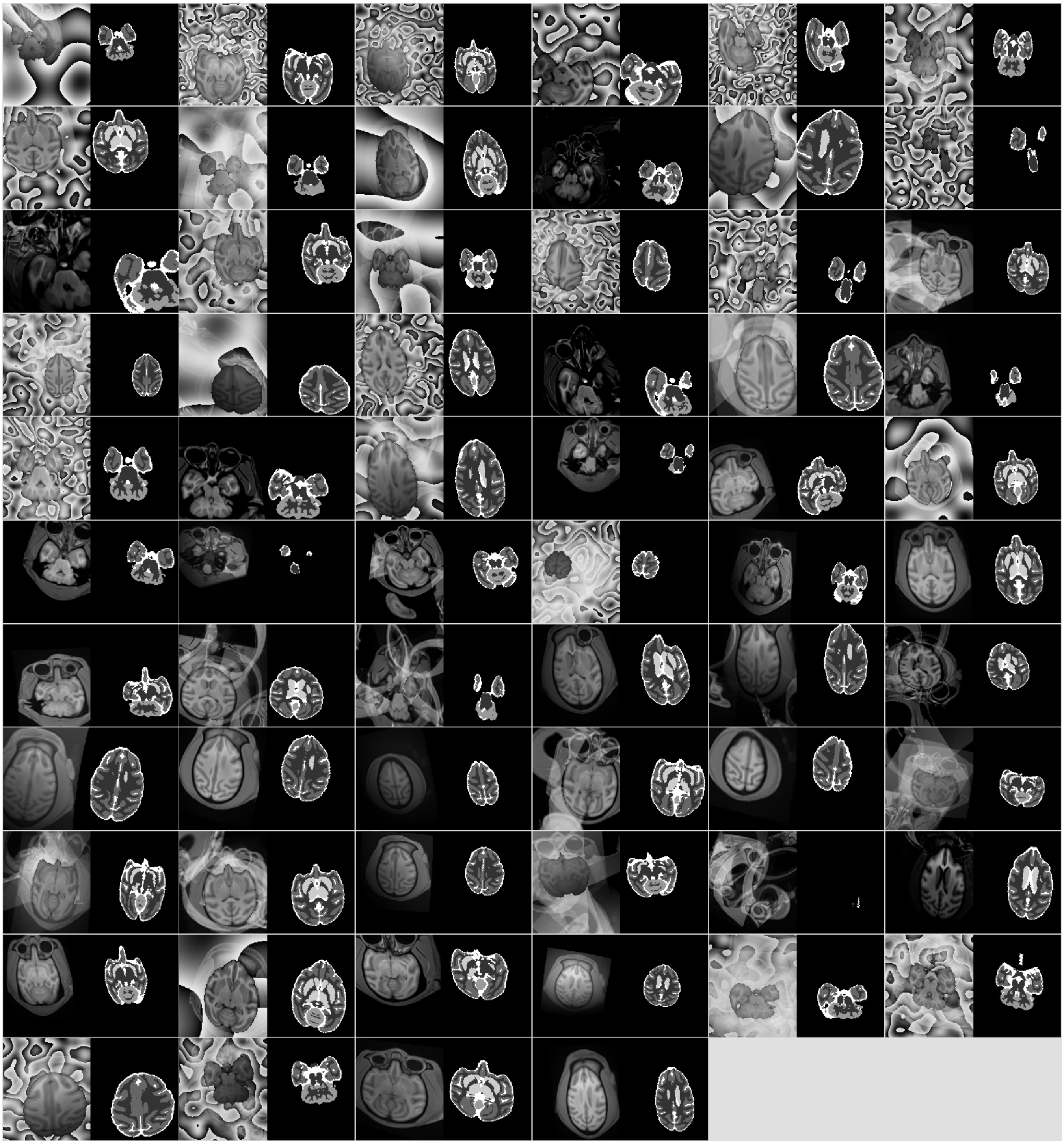

**Suppl Fig. 1**

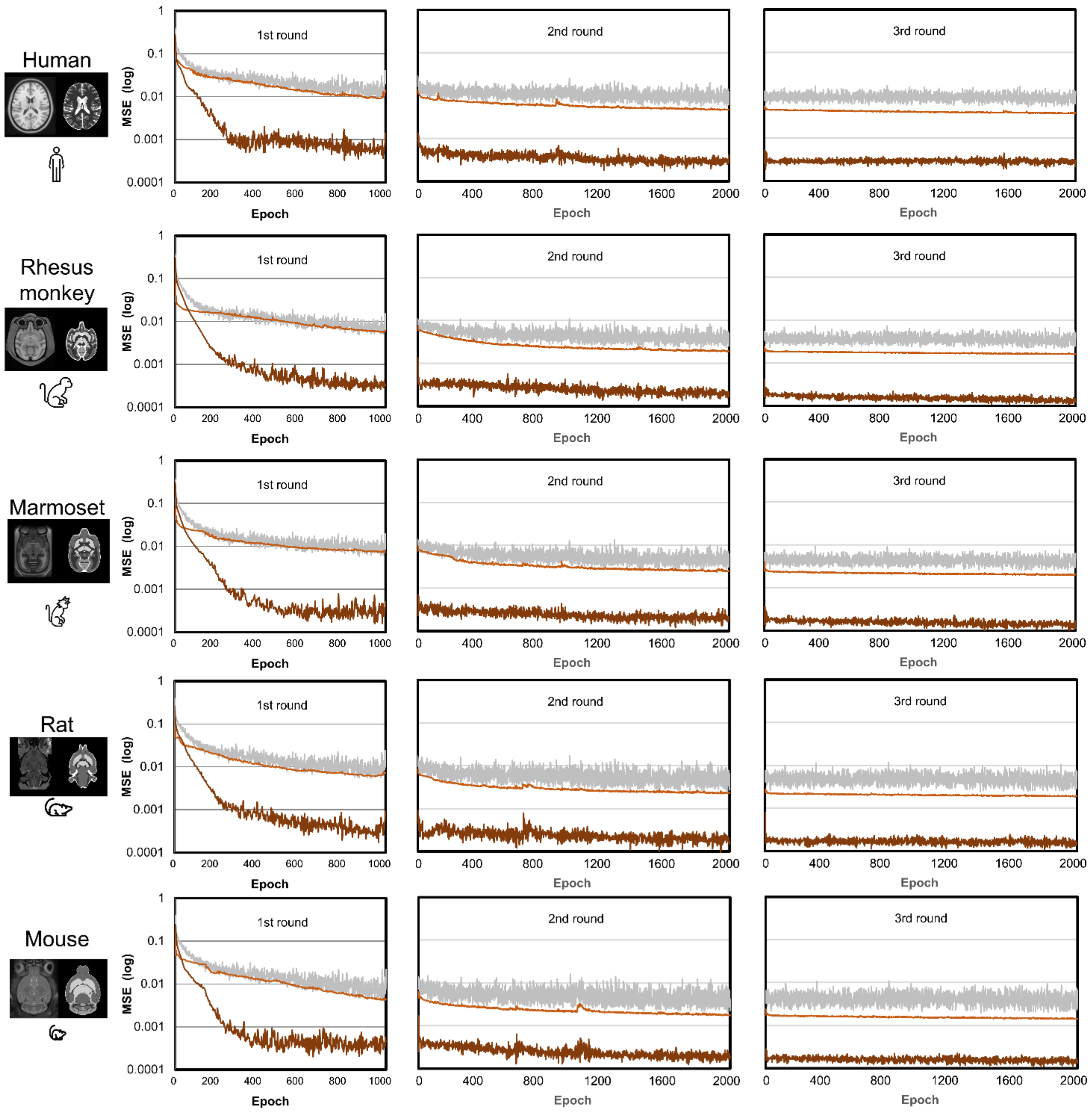

**Suppl Fig. 2**

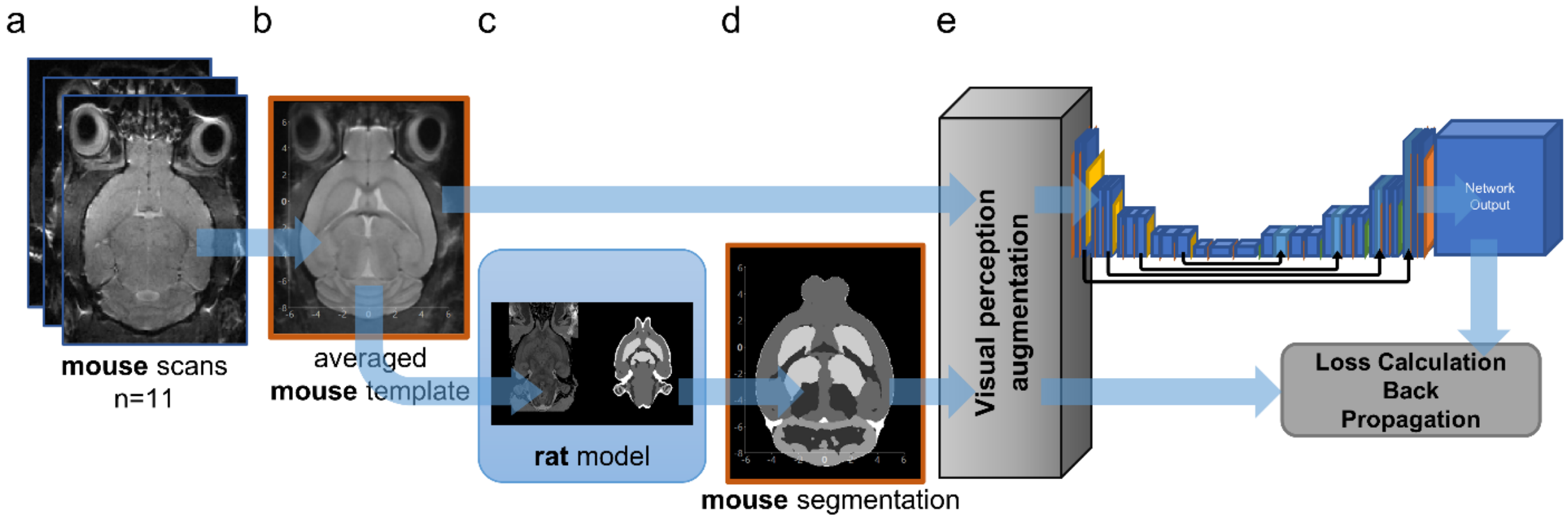

**Suppl Fig. 3**

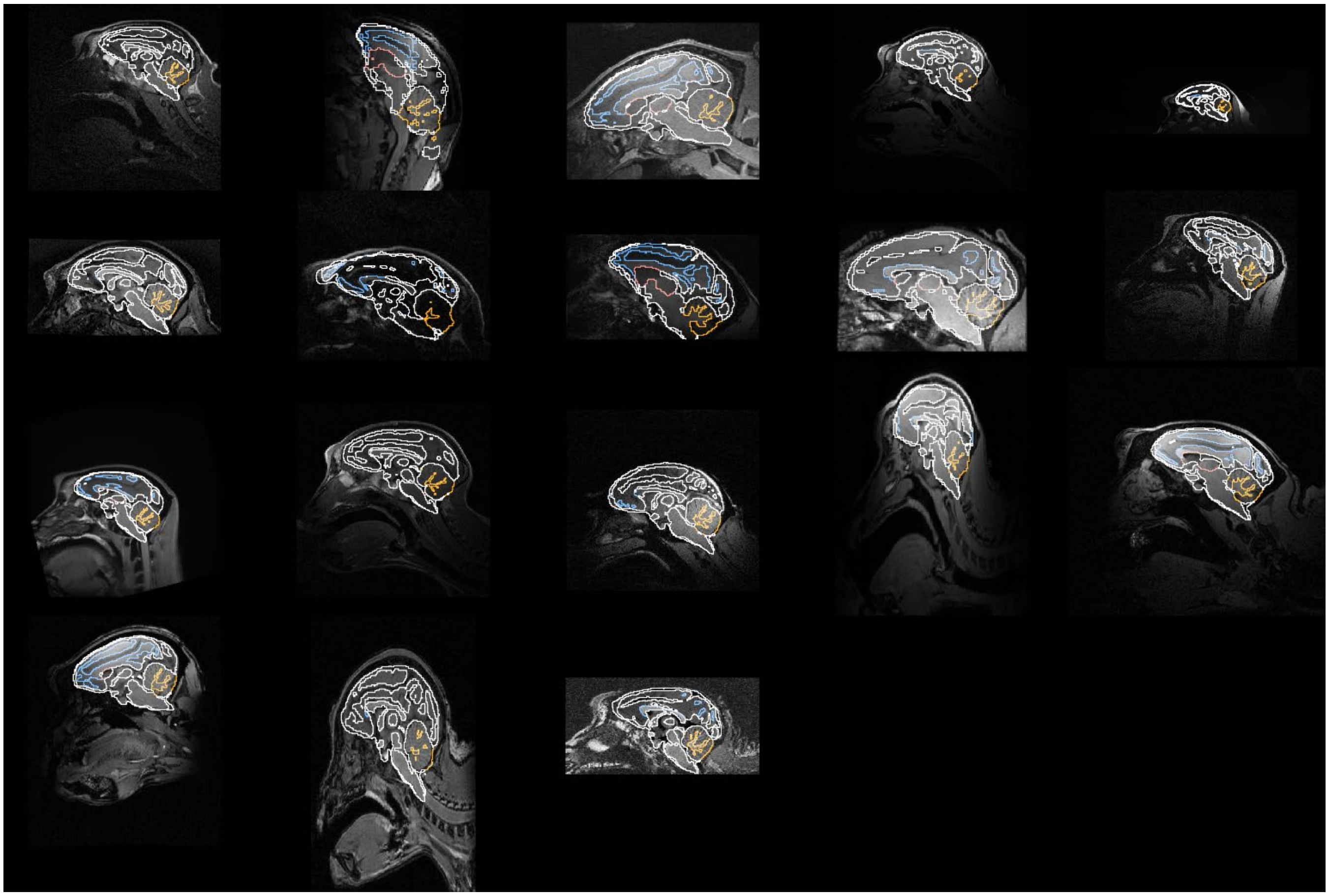

**Suppl Fig. 4**

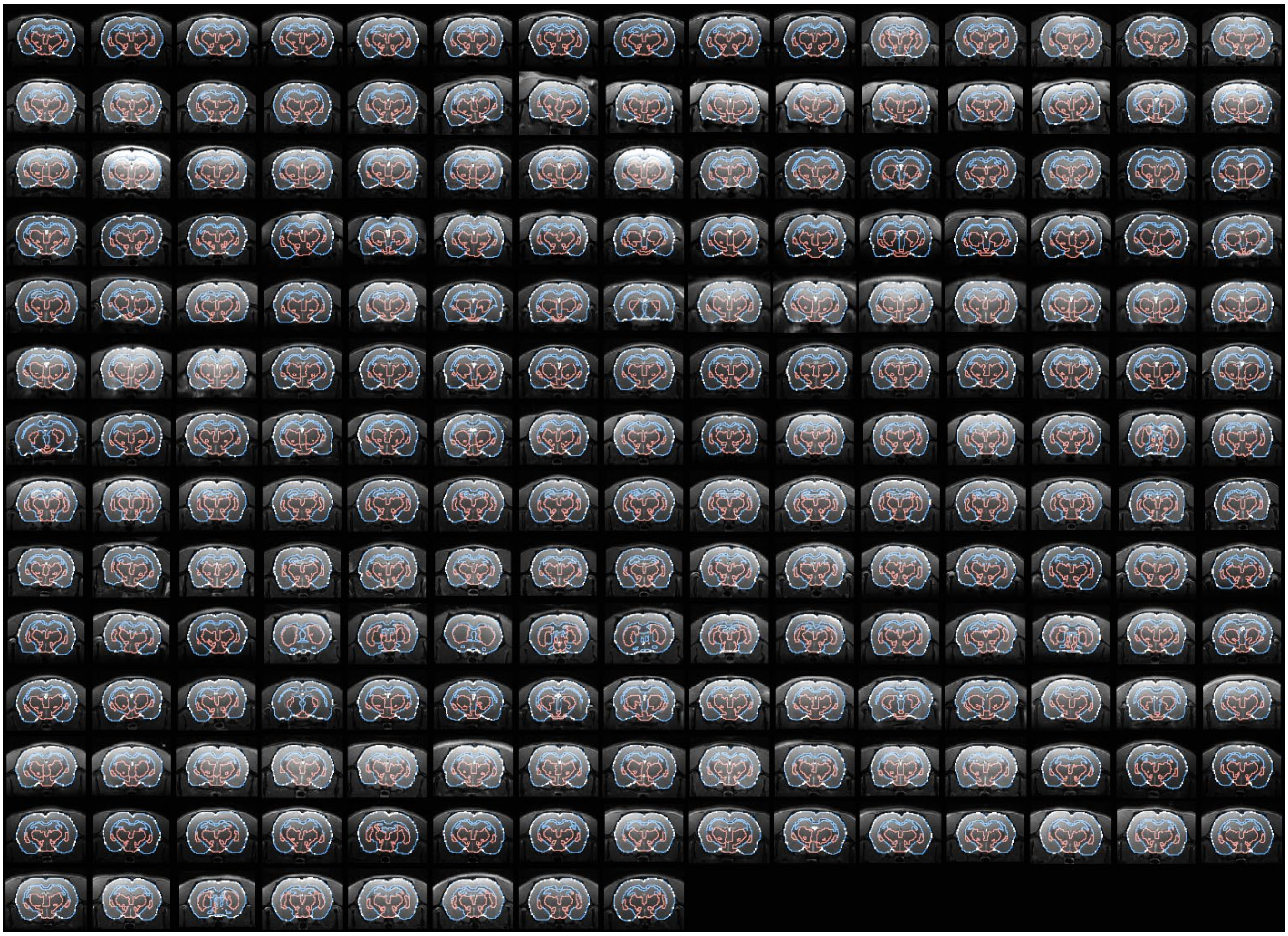

**Suppl Fig. 5**

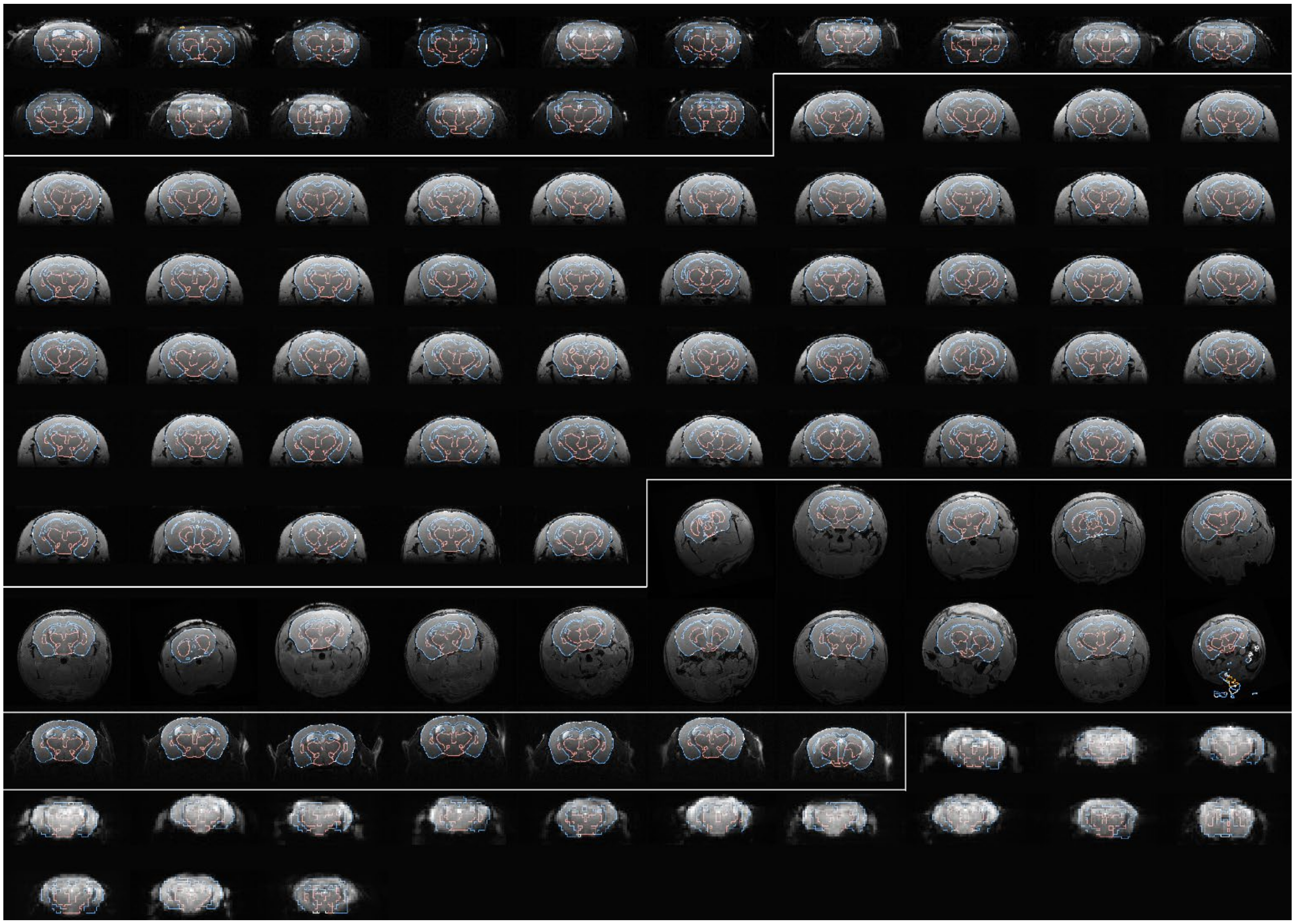

**Suppl Fig. 6**

# Supplementary Tables

**Supplementary Table 1: Source of Template Data for Visual Perception Augmentation**

| Species | Image Source | Label Source | Processing Applied | Label editing |
|---|---|---|---|---|
| **Human** | ICBM152 2009a T1W T2W [1]<br><br>https://www.bic.mni.mcgill.ca/ServicesAtlases/ICBM152NLin2009 | Freesurfer applied to ICBM152 2009a T1W images | The template Image was padded to 192×224×192 | Regions near brainstem, optic chiasm, and mid brain were edited on DSI Studio. |
| **Rhesus** | ONPRC18 Large FOV T1W T2W [2]<br><br>https://www.nitrc.org/projects/onprc18_atlas | ONPRC18 gray matter and while matter label maps.<br><br>Regions at cerebellum were edited on DSI Studio. | The template Image was padded to 192×224×192<br><br>The large FOV images were corrected by replacing images within the brain mask with the skull-stripped T1W T2W. | Regions near cerebellum were edited on DSI Studio. |
| **Marmoset** | Marmoset Brain Atlas V3 T1W T2W<br><br>https://marmosetbrainmapping.org/data.html | Marmoset Brain Atlas V3 T1W T2W | The template Image was padded to 192×224×192 | No additional editing applied |
| **Rat** | The MRI data of the SIGMA rat brain template<br><br>https://www.nitrc.org/projects/sigma_template | The MRI data of the SIGMA rat brain template | The template Image was padded to 260×342×184 | Regions near cerebellum were edited on DSI Studio. |
| **Mouse** | T2W images were nonlinearly averaged from Brookhaven National Laboratory (BNL) research<br><br>https://www.nitrc.org/projects/c57bl_mr_atlas | Generated using the rat SIGMA model | The T2W images were nonlinearly averaged from the n=11 BNL data<br><br>The template Image was padded to 288×352×224 | No additional editing applied |

[1] VS Fonov, AC Evans, K Botteron, CR Almli, RC McKinstry, DL Collins and BDCG, Unbiased average age-appropriate atlases for pediatric studies, NeuroImage,Volume 54, Issue 1, January 2011, ISSN 1053–8119, DOI: 10.1016/j.neuroimage.2010.07.033
[2] Weiss, Alison R., et al. "The macaque brain ONPRC18 template with combined gray and white matter labelmap for multimodal neuroimaging studies of nonhuman primates." Neuroimage 225 (2021): 117517.
[3] Liu C, et al. Marmoset Brain Mapping V3: Population multi-modal standard volumetric and surface-based templates. Neuroimage (2021) doi:10.1016/j.neuroimage.2020.117620.
[4] Barrière DA, Magalhães R, Novais A, Marques P, Selingue E, Geffroy F, Marques F, Cerqueira J, Sousa JC, Boumezbeur F, Bottlaender M. The SIGMA rat brain templates and atlases for multimodal MRI data analysis and visualization. Nature communications. 2019 Dec 13;10(1):5699.
[5] Ma, Y., et al., In Vivo 3D Digital Atlas Database of the Adult C57BL/6J Mouse Brain by Magnetic Resonance Microscopy. Front Neuroanat, 2008. 2: p. 1.

**Supplementary Table 2: Source of evaluation images**

| Species | Source | Image Information | Scan ID | Links |
|---|---|---|---|---|
| **Human** | Human Connectome Project Young Adult [1] | T1W preprocessed image at 0.75-mm isotropic resampled to 1-mm isotropic | 100206 | https://db.humanconnectome.org/ |
| **Chimpanzee** | National Chimpanzee Brain Resource | T1W image at 0.5-mm isotropic resolution | Agatha | https://www.chimpanzeebrain.org/ |
| **Rhesus** | PRIMatE Data Exchange (PRIME-DE) Mount Sinai School of Medicine (Philips) [2] | T1W image at 0.5-mm isotropic resolution | 032146 | https://fcon_1000.projects.nitrc.org/indi/indiPRIME.html |
| **Marmoset** | Brain/MINDS Marmoset Brain MRI Dataset NA216 and eNA91 [3] | T1W at 0.27-mm isotropic resolution | 001 | https://dataportal.brainminds.jp/marmoset-mri-na216 |
| **Rat** | Standard Rat [4] | T2-RARE at 0.2-mm isotropic resolution | 105 | https://openneuro.org/datasets/ds004116/versions/1.0.0 |
| **Mouse** | GDM offsprings [5] | T2w FLASH 3D at | K6M72 | https://openneuro.org/datasets/ds004145/versions/1.0.0 |

[1] Van Essen DC, Smith SM, Barch DM, Behrens TE, Yacoub E, Ugurbil K, Wu-Minn HCP Consortium. The WU-Minn human connectome project: an overview. Neuroimage. 2013 Oct 15;80:62-79.
[2] Milham M, Petkov CI, Margulies DS, Schroeder CE, Basso MA, Belin P, Fair DA, Fox A, Kastner S, Mars RB, Messinger A. Accelerating the evolution of nonhuman primate neuroimaging. Neuron. 2020 Feb 19;105(4):600-3.
[3] Hata J, Nakae K, Tsukada H, Woodward A, Haga Y, Iida M, Uematsu A, Seki F, Ichinohe N, Gong R, Kaneko T. Multi-modal brain magnetic resonance imaging database covering marmosets with a wide age range. Scientific Data. 2023 Apr 27;10(1):221.
[4] Grandjean J, Desrosiers-Gregoire G, Anckaerts C, Angeles-Valdez D, Ayad F, Barrière DA, Blockx I, Bortel A, Broadwater M, Cardoso BM, Célestine M. A consensus protocol for functional connectivity analysis in the rat brain. Nature neuroscience. 2023 Mar 27:1-9.
[5] Xin Yi Yeo and HanGyu Bae and Ling-Yun Yeow and Hongyu Li and Li Yang Tan and Woo Ri Chae and Joanes Grandjean and Weiping Han and Sangyong Jung (2022). GDMOffspring_MRI. OpenNeuro. [Dataset] doi: doi:10.18112/openneuro.ds004145.v1.0.0

**Supplementary Table 3: Blind evaluation results**

| Evaluator Votes (Sorted) | A versus B | | A versus C | | B versus C | | Evaluator background | |
|---|---|---|---|---|---|---|---|---|
| | A>B | B>A | A>C | C>A | B>C | C>B | PhD | MD |
| A=CB | 1 | | | | | 1 | x | |
| A=CB | 1 | | | | | 1 | | |
| AB=C | 1 | | 1 | | | | x | |
| AB=C | 1 | | 1 | | | | | x |
| AB=C | 1 | | 1 | | | | | x |
| ABC | 1 | | 1 | | 1 | | x | |
| ABC | 1 | | 1 | | 1 | | x | |
| ABC | 1 | | 1 | | 1 | | x | x |
| ACB | 1 | | 1 | | | 1 | x | |
| ACB | 1 | | 1 | | | 1 | x | |
| ACB | 1 | | 1 | | | 1 | x | |
| ACB/CBA | 0.5 | 0.5 | 0.5 | 0.5 | | 1 | | x |
| BCA | | 1 | | 1 | 1 | | | |
| BCA | | 1 | | 1 | 1 | | | x |
| CA=B | | | | 1 | | 1 | x | |
| CA=B | | | | 1 | | 1 | x | |
| CAB | 1 | | | 1 | | 1 | x | x |
| CAB | 1 | | | 1 | | 1 | x | |
| CAB | 1 | | | 1 | | 1 | x | |
| CAB | 1 | | | 1 | | 1 | x | |
| CAB | 1 | | | 1 | | 1 | | |
| CAB | 1 | | | 1 | | 1 | x | |
| CAB | 1 | | | 1 | | 1 | x | |
| CAB | 1 | | | 1 | | 1 | x | |
| CAB | 1 | | | 1 | | 1 | x | |
| CAB | 1 | | | 1 | | 1 | | |
| CAB | 1 | | | 1 | | 1 | x | |
| CAB | 1 | | | 1 | | 1 | | x |
| CAB | 1 | | | 1 | | 1 | | |
| CBA | | 1 | | 1 | | 1 | x | |
| CBA | | 1 | | 1 | | 1 | x | |
| CBA | | 1 | | 1 | | 1 | x | |
| CBA | | 1 | | 1 | | 1 | | x |
| CBA | | 1 | | 1 | | 1 | x | |
| CBA | | 1 | | 1 | | 1 | | |
| | 24.5 | 8.5 | 9.5 | 23.5 | 5 | 27 | 23 | 8 |